\documentclass{aastex}

\usepackage{graphicx}
\usepackage{longtable}
\usepackage{subfigure}

\newcommand{\sn}{SN 2014cx}
\newcommand{\host}{NGC 337}
\newcommand{\ha} {\mbox{H$\alpha$}}
\newcommand{\hb} {\mbox{H$\beta$}}

\newcommand{\ld}{\mbox{$\lambda$}}
\newcommand{\eg}{{\textrm e.g.,}}
\newcommand{\ie}{{\textrm i.e.,}}
\newcommand{\msun}{\mbox{M$_{\odot}$}}
\newcommand{\rsun}{\mbox{R$_{\odot}$}}
\newcommand{\kms}{\mbox{$\rm{\,km\,s^{-1}}$}}
\newcommand{\ergs}{\mbox{$\rm{\,erg\,s^{-1}}$}}

\newcommand{\nickel}{\mbox{$^{56}$Ni}}
\newcommand{\cobalt}{\mbox{$^{56}$Co}}
\newcommand{\iron}{\mbox{$^{56}$Fe}}

\shorttitle{Type IIP Supernova 2014cx}
\shortauthors{Huang et al.}

\begin{document}

\title{Optical and Ultraviolet Observations of the Very Young Type IIP \sn\ in \host }

\author{
Fang Huang\altaffilmark{1},
Xiaofeng Wang\altaffilmark{1},
Luca Zampieri\altaffilmark{2},
Maria Letizia Pumo\altaffilmark{3},
Iair Arcavi\altaffilmark{4,5},
Peter J. Brown\altaffilmark{6},
Melissa L. Graham\altaffilmark{7},
Alexei V. Filippenko\altaffilmark{7},
WeiKang Zheng\altaffilmark{7},
Griffin Hosseinzadeh\altaffilmark{4},
D. Andrew Howell\altaffilmark{4,8},
Curtis McCully\altaffilmark{4},
Liming Rui\altaffilmark{1},
Stefano Valenti\altaffilmark{9},
Tianmeng Zhang\altaffilmark{10},
Jujia Zhang\altaffilmark{11,12},
Kaicheng Zhang\altaffilmark{1},
Lifan Wang\altaffilmark{6}}

\altaffiltext{1} {Physics Department and Tsinghua Center for Astrophysics, Tsinghua University, Beijing, 100084, China;  \mbox{huangfang@mail.tsinghua.edu.cn; wang\_xf@mail.tsinghua.edu.cn}}
\altaffiltext{2} {INAF-Osservatorio Astronomico di Padova, Vicolo dell'Osservatorio 5, 35122 Padova, Italy}
\altaffiltext{3} {INAF-Osservatorio Astronomico di Palermo ``Giuseppe S. Vaiana," Piazza del Parlamento 1, 90134 Palermo, Italy}
\altaffiltext{4} {Las Cumbres Observatory Global Telescope Network, 6740 Cortona Dr., Suite 102, Goleta, CA 93117, USA}
\altaffiltext{5} {Kavli Institute for Theoretical Physics, Kohn Hall, University of California, Santa Barbara, CA 93106-4030, USA}
\altaffiltext{6} {George P. and Cynthia Woods Mitchell Institute for Fundamental Physics \& Astronomy, Texas A\&M University, Department of Physics and Astronomy, 4242 TAMU, College Station, TX 77843, USA}
\altaffiltext{7} {Department of Astronomy, University of California, Berkeley, CA 94720-3411, USA}
\altaffiltext{8} {Department of Physics, University of California, Santa Barbara, Broida Hall, Mail Code 9530, Santa Barbara, CA 93106-9530, USA}
\altaffiltext{9} {Department of Physics, University of California, Davis, CA 95616, USA}
\altaffiltext{10} {Key Laboratory of Optical Astronomy, National Astronomical Observatories, Chinese Academy of Sciences, Beijing 100012, China}
\altaffiltext{11} {Yunnan Observatories, Chinese Academy of Sciences, Kunming 650011, China}
\altaffiltext{12} {Key Laboratory for the Structure and Evolution of Celestial Objects, Chinese Academy of Sciences, Kunming 650216, China}

\begin{abstract}
Extensive photometric and spectroscopic observations are presented for \sn, a type IIP supernova (SN) exploding in the nearby galaxy \host. The observations are performed in optical and ultraviolet bands, covering from $-20$ to +400 days from the peak light. The stringent detection limit from prediscovery images suggests that this supernova was actually detected within about 1 day after explosion. Evolution of the very early-time light curve of \sn\ is similar to that predicted from a shock breakout and post-shock cooling decline before reaching the optical peak. Our photometric observations show that SN 2014cx has a plateau duration of $\sim 100$ days, an absolute $V$-band magnitude of $\sim -16.5$ mag at $t \approx 50$ days, and a nickel mass of $0.056 \pm 0.008$ M$_{\odot}$. The spectral evolution of SN 2014cx resembles that of normal SNe~IIP like SN 1999em and SN 2004et, except that it has a slightly higher expansion velocity ($\sim 4200$ \kms\ at 50 days). From the cooling curve of photospheric temperature, we derive that the progenitor has a pre-explosion radius of $\sim 640$ \rsun, consistent with those obtained from SNEC modeling ($\sim 620$ \rsun) and hydrodynamical modeling of the observables ($\sim 570$ \rsun). Moreover, the hydrodynamical simulations yield a total explosion energy of $\sim 0.4 \times 10^{51}$ erg, and an ejected mass of $\sim 8$ M$_{\odot}$. These results indicate that the immediate progenitor of \sn\ is likely a red supergiant star with a mass of $\sim 10$ M$_{\odot}$.

\end{abstract}

\keywords{supernovae: general $-$ supernovae: individual: {\sn} $-$ galaxies: individual: \host}

\section{Introduction} \label{sec:intro}
Type IIP supernovae (SNe IIP) represent the most common subtype of stellar explosions, constituting about one third of all SNe \citep{2011MNRAS.412.1441L}. This subtype of SNe is thought to arise from core-collapse (CC) explosions of massive red supergiants (RSGs) with an  initial mass of 8--25 \msun\ according to theoretical models of stellar evolution \citep{2003ApJ...591..288H}. On the other hand, direct analyses of supernova position on pre-explosion images give a much narrower range for the progenitor mass, e.g., 8.5--16.5 \msun\ \citep{2007ApJ...661.1013L, 2009ARA&A..47...63S}. Compared to other CC SNe, SNe IIP are characterized by prominent hydrogen features in their optical spectra \citep[e.g.,][]{1997ARA&A..35..309F} and an extended plateau phase in their light curves. During the plateau phase, their luminosity remains almost constant as a result of the energy balance between the hydrogen recombination and expansion cooling. The plateau feature distinguishes SNe~IIP from the Type IIL subclass, for which the light curve exhibits a linear decline (in mag day$^{-1}$) after the peak \citep{1979A&A....72..287B}. Recent statistical analyses show that the light-curve properties of SNe II may have a continuous distribution (\eg\ \citealt{2014ApJ...786...67A, 2015ApJ...799..208S, 2016MNRAS.459.3939V}), although there are also studies suggesting a distinct division between type IIP and IIL SNe \citep{2012ApJ...756L..30A, 2014MNRAS.442..844F, 2014MNRAS.445..554F}.

Over the years, numerous SNe IIP have been well studied, such as SN 1999em \citep{2002PASP..114...35L}, SN 2004et \citep{2006MNRAS.372.1315S}, SN 2005cs \citep{2009MNRAS.394.2266P}, and SN 2013ej \citep{2015ApJ...807...59H}. These studies reveal a large spread in luminosities, plateau durations, expansion velocities, and nickel masses for SNe IIP (\eg\ \citealt{2003ApJ...582..905H}), which can be well understood with current explosion models (\eg\ \citealt{2009ApJ...703.2205K, 2010MNRAS.408..827D, 2011ApJ...741...41P, 2013MNRAS.434.3445P}). Nevertheless, early time observations are still sparse for SNe IIP, which are vital to constrain the explosion time and hence determine the properties of their progenitor stars \citep{2003MNRAS.346...97N, 2009MNRAS.395.1409S}. In particular, very early light curves of SNe IIP may be affected by a short, sharp blast of light as a result of shock breakout of the stellar surface, as predicted in core collapse explosion of massive stars \citep{1977ApJS...33..515F, 1978ApJ...223L.109K}. \sn\ represents such a CC SN that is captured within about one day after the explosion.

\sn\ was independently discovered on UT 2014 Sept. 2 by \citet{2014CBET.3963....1N} and \citet{2014ATel.6436....1H} in the nearby SBd galaxy \host. Based on the astrometry from the USNO-A2.0 Catalogue \citep{1998usno.book.....M}, the J2000 coordinates of the SN are derived as $\alpha$ = 00$^h$59$^m$47.83$^s$ and $\delta = -07\arcdeg 34 \arcmin 19.3 \arcsec$, approximately 21.7$\arcsec\,$N and 33.7$\arcsec\,$W from the center of \host\ \citep{2014ATel.6436....1H}. \sn\ was reported as a young SN II based on both optical \citep{2014ATel.6440....1E} and near-infrared (NIR; \citealt{2014ATel.6442....1M}) spectra taken at about one day after the discovery. It was further classified as a Type IIP event according to the photometric observations by \cite{2015ATel.7084....1A}. We note that another SN~IIP, SN 2011dq, also exploded in \host. The distance to \host\ is estimated to be $18.0\pm3.6$ Mpc (distance modulus $\mu=31.27\pm0.43$ mag) by the Tully-Fisher method \citep{2014MNRAS.444..527S}; here we adopt this value for \sn.

In this work, we present the results of our optical and UV observations of the type IIP supernova \sn\ that was discovered at a very young age. The observations and data reduction are addressed in Section \ref{sec:obs}, the photometric and spectroscopic evolution are described in Section \ref{sec:lc} and \ref{sec:spec}, respectively, and  analysis of the progenitor properties of \sn\ via the photospheric temperature cooling curve and hydrodynamical modeling is given in Section \ref{sec:model}. The main results are summarized in Section \ref{sec:sum}.

\section{Observations and Data Reduction} \label{sec:obs}
\subsection{Photometry} \label{sec:obs.phot}
\subsubsection{Optical Observations}
We started the $UBVRI$ follow-up campaign with the 0.8~m Tsinghua University-NAOC telescope (hereafter TNT) at Xinglong Observatory in China \citep{2008ApJ...675..626W, 2012RAA....12.1585H}. The TNT monitoring of \sn\ started on 2014 Sep. 15 and continued until 2015 Jan. 31. High-cadence Johnson $BV$ and Sloan $gri$ monitoring was conducted using the 1.0~m telescopes in the Las Cumbres Observatory Global Telescope network (hereafter LCOGT; \citealt{2013PASP..125.1031B}). The LCOGT data cover 2014 Sep. 3 through 2015 Sep. 17. In addition, 55 epochs of unfiltered data were collected with the 0.76~m Katzman Automatic Imaging Telescope (KAIT; \citealt{2001ASPC..246..121F}), extending to 2015 Jan. 5.

\begin{figure}[!hbp]
\centering
\includegraphics*[width=0.8\textwidth,trim=500 140 540 100]{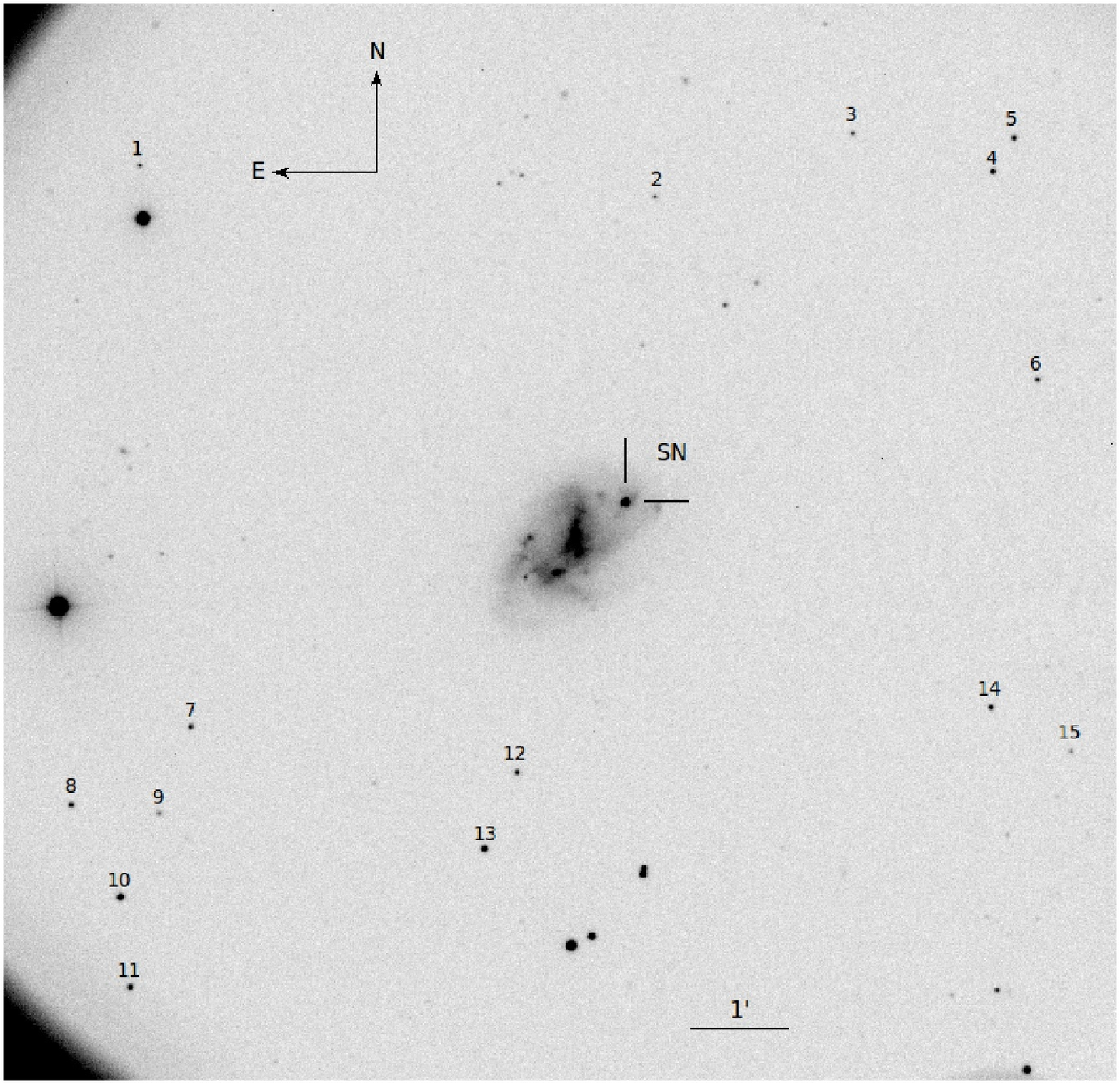}
\caption{\sn\ in \host. This $R$-band image was taken on 2014 Oct. 25 with the 80~cm Tsinghua-NAOC telescope. \sn\ and the 15 local sequence stars are marked.}
\label{fig:finder}
\end{figure}

All of the images were reduced with standard \textsc{IRAF}\footnote{IRAF is distributed by the National Optical Astronomy Observatories, which are operated by the Association of Universities for Research in Astronomy, Inc., under cooperative agreement with the National Science Foundation (NSF).} routines. Point-spread function (PSF) photometry was performed on TNT and LCOGT data using the \textsc{SNOoPy} package\footnote{http://sngroup.oapd.inaf.it/snoopy.html}. KAIT data were reduced using a PSF-based image-reduction pipeline \citep{2010ApJS..190..418G}. The SN instrumental magnitudes were calibrated using 15 field stars (marked in Figure \ref{fig:finder}) from the Sloan Digital Sky Survey (SDSS) Data Release 9 catalog \citep{2012ApJS..203...21A} and transformed to the Johnson system. The magnitudes of the reference stars are listed in Table \ref{tab:standstar}. The final flux-calibrated optical magnitudes of \sn\ are shown in Table \ref{tab:tnt}--\ref{tab:kait}.

\subsubsection{Swift Ultraviolet Observations}
In addition to the ground-based observations, \sn\ was also monitored with the Ultraviolet/Optical Telescope (UVOT; \citealt{2005SSRv..120...95R}) on board the \emph{Swift} spacecraft through the $uvw2, uvm2, uvw1,u, b$, and $v$ filters. The UV observations began on 2014 Sep. 3 and ended on 2014 Oct. 24. The data were obtained from the \emph{Swift} Optical/Ultraviolet Supernova Archive (SOUSA; \citealt{2014Ap&SS.354...89B}). The reduction method for UVOT photometry is based on that of \citet{2009AJ....137.4517B}, which includes aperture photometry after subtracting off the underlying galaxy count rates, and adopting the updated zeropoints and time-dependent sensitivity from \citet{2011AIPC.1358..373B}. The final UVOT photometry of \sn\ is presented in Table \ref{tab:uvot}.

\subsection{Spectroscopy} \label{sec:obs.spec}
The spectroscopic monitoring campaign of \sn\ started on 2014 Sep. 7 and continued for $\sim 400$ days. Fourteen low-resolution optical spectra were obtained using the LCOGT 2~m Faulkes Telescope South (FTS; with FLOYDS), the Xinglong 2.16~m telescope (with BFOSC), the Lijiang 2.4~m telescope (with YFOSC), the LCOGT 2~m Faulkes Telescope North (FTN; with FLOYDS), the GEMINI North Telescope (GNT; with GMOS), and the Keck-I 10~m telescope (with LRIS); see Table \ref{tab:spelog} for the detailed information on the spectroscopic observations.

The spectra were all reduced in a standard manner using various tasks within \textsc{IRAF}. After the preliminary reduction (including bias/overscan correction, flat fielding, and cosmic-ray removal), the one-dimensional spectra were extracted using the optimal extraction method \citep{1986PASP...98..609H}. Wavelengths of the SN spectra were calibrated using the Fe/Ar and Fe/Ne lamp spectra. For the FLOYDS spectra, a Hg/Ar lamp is used for wavelength calibration. The spectra were then flux calibrated using the instrumental sensitivity curves of spectrophotometric standard stars observed on the same (or nearby) night and with the same instrumental setup.

\section{Photometric Analysis}  \label{sec:lc}

\subsection{The Early-Time Evolution} \label{subsec:lc}
With modern high-cadence surveys, we are able to detect young SNe and trigger follow-up observations immediately. Such early-time detections help constrain the explosion properties and (with luck) even reveal shock breakout, which often happens hours after explosion \citep{2007ApJ...666.1093Q, 2008Sci...321..223S, 2009ApJ...705L..10T, 2015ApJ...804...28G, 2016ApJ...820...23G, 2016ApJ...822....6D}.

\begin{figure}
\centering
\includegraphics[width=0.8\textwidth]{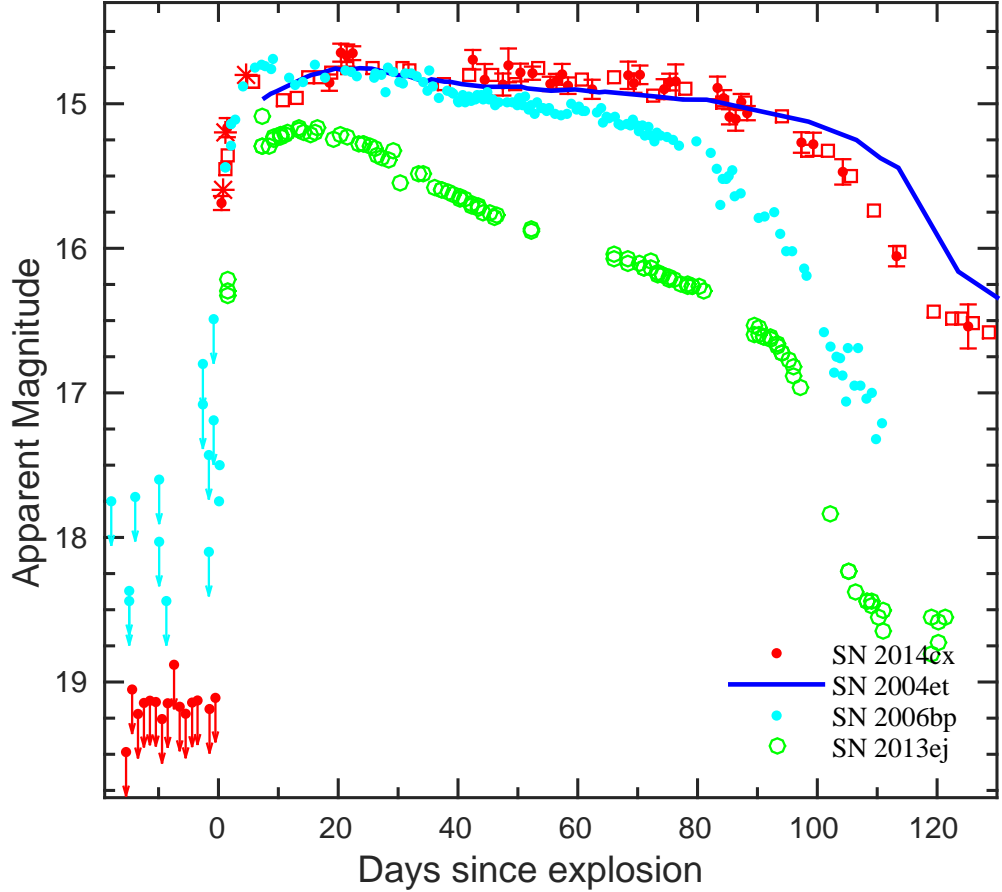}
\caption{KAIT unfiltered (red circle) and LCOGT $r$ band (red square) light curve of \sn\, unfiltered detections from the \textit{Bright Supernovae} website (http://www.rochesterastronomy.org/snimages/) (red star), SN 2004et in the $R$ band (blue line), SN 2006bp (cyan), and SN 2013ej (green). For KAIT data, detections are plotted with filled circles, and arrows represent 4$\sigma$ upper limits.}
\label{fig:lc_kait}
\end{figure}

The field of SN 2014cx was monitored daily by KAIT in the clear band (\ie\, unfiltered) before the explosion, leading to the earliest detection of this object after the explosion. Figure \ref{fig:lc_kait} shows the KAIT unfiltered light curve, spanning from a few weeks before the explosion to $\sim 120$ days after. A few early-time observations obtained by amateurs in unfiltered mode are also overplotted. As can be seen, \sn\ was not detected on 2014 Sep. 1 (MJD = 56,901.39) with a limit of $> 19.1$ mag, but it was detected one day later (MJD = 56,902.40) at 15.69 mag. We therefore adopt this epoch (MJD=56901.89 $\pm$ 0.5) as the reference date for the shock breakout for \sn, and this determination sets one of the tightest constraints on the shock breakout time for a SN IIP based on the observations of ground-based telescopes.

The unfiltered light curve of \sn\ at early time (0--20 days) is very similar to those of SN 2006bp and SN 2013ej. It seems to experience three stages. First, the SN brightened by more than 4 mag over the first 1.5--2.0 days, and it then dimmed and rebrightened toward the primary peak at $t \approx 20$ days. This trend is actually more prominent in the early-time evolution of the LCOGT $r$-band light curve (see Figure \ref{fig:lc_kait}), with a noticeable dip occurring at t $\approx$ 10 days after explosion. Such a behavior is also seen in the comparison object SN 2006bp (and possibly SN 2013ej), which is likely related to the shock breakout of the supernova surface and its cooling phase \citep{2007ApJ...666.1093Q}.

\subsection{Overall Evolution of the Light Curves}
\begin{figure}
\centering
\includegraphics[width=0.8\textwidth]{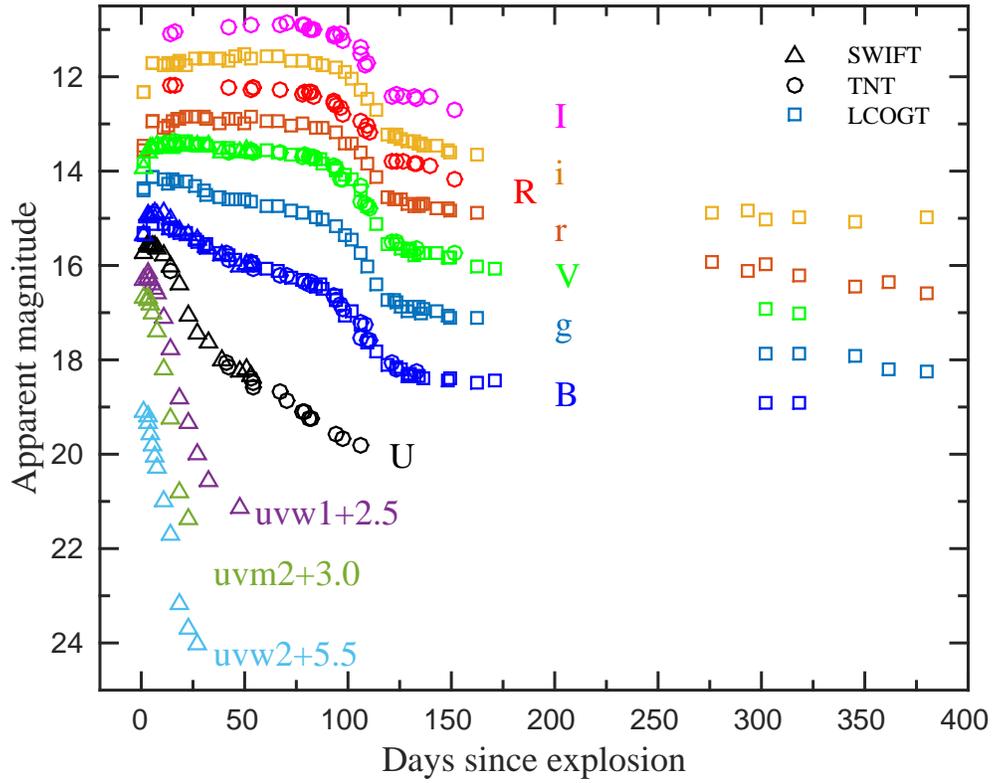}
\caption{The photometric evolution of \sn\ in UV and optical bands. The phase is given relative to the estimated explosion date, MJD = 56,901.89.}  \label{fig:lc}
\end{figure}

Figure \ref{fig:lc} shows the multicolor photometric evolution of \sn\ during the period from $t= +1$ to $t=+380$ days after the explosion. At the earliest phases, the light curves exhibit a rapid rise in the UV and optical bands, but with a slower pace at longer wavelengths (see Table \ref{tab:photopara}). By fitting a low-order polynomial to the data around maximum light, we determine the magnitudes and corresponding dates at peak brightness. These results are also presented in Table \ref{tab:photopara}. We note that the rise time in the $B$ band is $\sim 8$ days, which agrees well with the statistical result from a large sample of SNe II \citep{2015MNRAS.451.2212G}.

One can see from Figure \ref{fig:lc} that the UV light curves show rapid post-peak declines, with slopes of 0.16, 0.23, and 0.25 mag d$^{-1}$ in the $uvw1, uvw2$, and $uvm2$ bands, respectively. The optical light curves, in contrast, drop slowly and settle to the plateau phase. A decline rate of ${\beta}^B_{100} = 2.5$ mag is measured for the $B$-band light curve over the first 100 d after maximum light, consistent with the typical value obtained for normal SNe IIP (i.e., ${\beta}^B_{100} < 3.5$ mag; \citealt{1994A&A...282..731P}). The $V$-band brightness declines by $\sim$ 0.2 mag in the first 50 days after the peak, which also falls into the SN~IIP group according to the criterion of \citet{2014MNRAS.445..554F}. For SN 2014cx, the plateau phase lasts for $\sim100$ days, and its brightness stays roughly constant, especially in the $VrRiI$ bands.

After the plateau, the \sn\ experiences a sudden drop in brightness in all optical bands, with magnitude declines of 1.5--2.0 mag in $\sim$ 20 days. To derive the parameters about the transitional phase, we fit the $V$-band light curve using Eq.(4) from \citet{2010ApJ...715..833O}. The middle date of the transition phase, $t_{\rm PT}$, is found to be $\sim 109$ d, and the width of the transition phase is $\sim 7$ d. During the tail phase (i.e., $t \approx 130$--170 days), the decline rates in $B$, $V$, $R$, and $I$ are measured to be 0.42, 0.95, 1.09, and 1.13  \mbox{mag (100 d)$ ^{-1}$}, respectively. These values are similar to those of normal SNe IIP such as SN 1999em \citep{2003MNRAS.338..939E}. In the $V$ and $R$ bands, the decline rates are found to be comparable to the values expected for the radioactive decay, 0.98 \mbox{mag (100 d)$ ^{-1} $} \citep{1994A&A...282..731P}. This indicates that the gamma-ray photons were effectively trapped in the ejecta at this phase. After $t \approx 170$ days, the light curves tend to show slower decline rates relative to the earlier nebular phase. For example, the decline rate of $B$- and $V$-band light curve is measured to be 0.32 and 0.65 \mbox{mag (100 d)$ ^{-1}$} during the period from 170 days $<$ t $<$ 320 days. And this decay rate is found to be 0.46, 0.85, and 0.81 \mbox{mag (100 d)$ ^{-1}$} in $gri$ bands at similar phase. This slower evolution can be due to interaction of SN ejecta with the circumstellar materials surrounding the SN or a scattered light echo (see discussions in \S 4.1).

\subsection{Color Curves} \label{subsec:colorlc}
The Galactic reddening in the direction of \host\ is $E(B-V)_{\rm MW} = 0.10$ mag \citep{2011ApJ...737..103S}. We didn't detect any significant features of Na~I~D absorption produced by the host galaxy in our 14 low-resolution spectra, suggesting that host-galaxy reddening might be negligible for \sn. As an alternative, we also use the color method proposed by \citet{2010ApJ...715..833O} to estimate the reddening due to the host galaxy. This method assumes that the intrinsic $V-I$ color of SNe IIP is constant (i.e., $(V-I)_0$=0.656 mag) toward the end of the plateau. After correcting for the Galactic reddening, the $V-I$ color of \sn\ is found to be 0.596 mag at $t = 79$ days, which gives a negative host-galaxy extinction for this object. Thus, we adopt $E(B-V)_{\rm tot} = 0.10$ mag in the following analysis, which corresponds to a total line-of-sight extinction of $A_V$ = 0.31 mag assuming the ratio of total to selective extinction $R_V$ = 3.1 \citep{1989ApJ...345..245C}.

\begin{figure}
\centering
\includegraphics[width=0.8\textwidth]{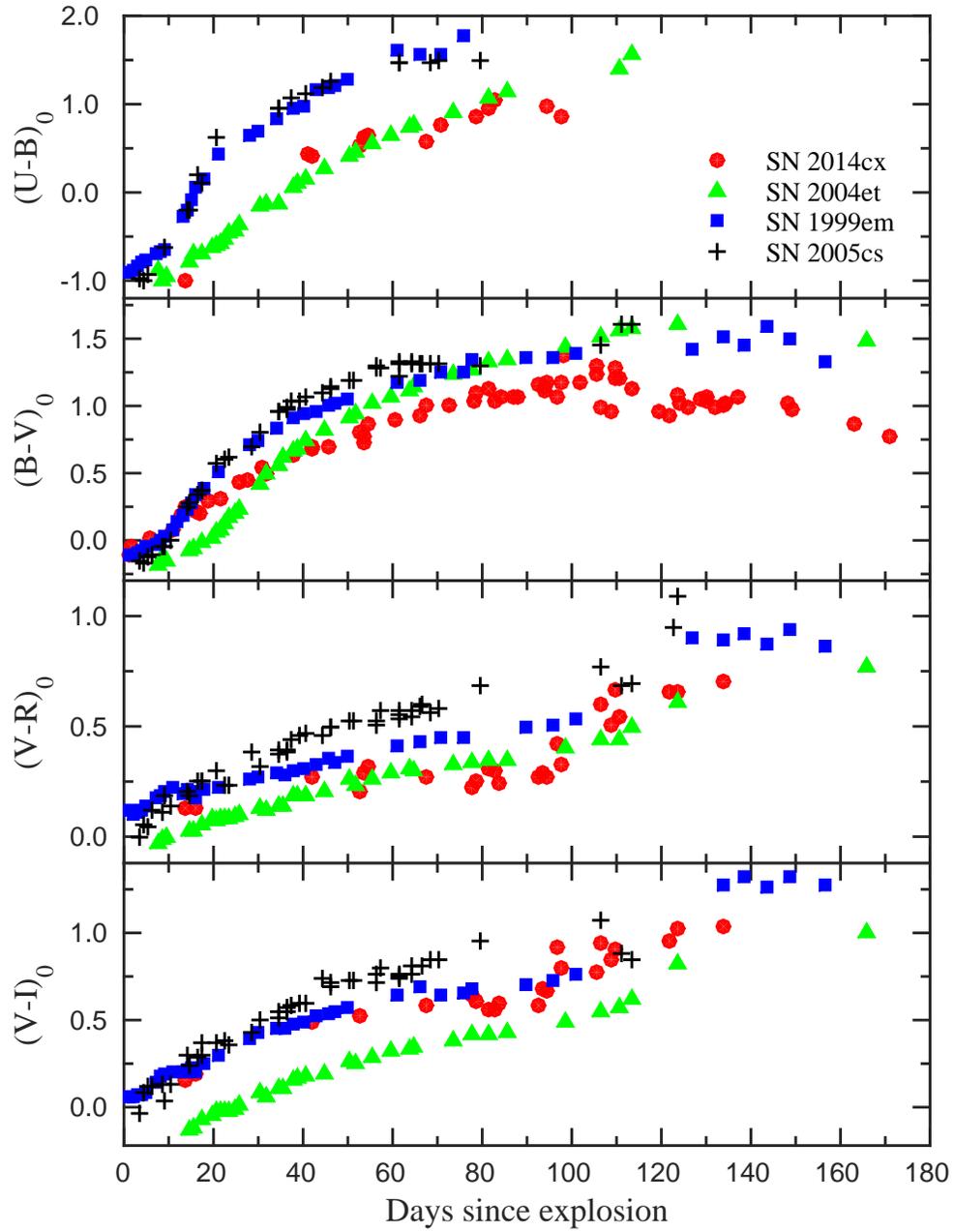}
\caption{The dereddening optical color evolution of \sn\ compared with those of other SNe IIP : SNe 1999em, 2004et, and 2005cs.}  \label{fig:color}
\end{figure}

The reddening-corrected $(U-B)_0$, $(B-V)_0$, $(V-R)_0$, and $(V-I)_0$ color curves of \sn\ are shown in Figure \ref{fig:color}. For comparison, we overplot the dereddened color curves of three well-studied type IIP SNe 1999em, 2004et, and 2005cs. The color evolution of \sn\ follows the general trend seen in SNe IIP: a rapid decrease from blue (high temperature) to red (low temperature) during the photospheric phase. The $(U-B)_0$ and $(B-V)_0$ colors become redder by $\sim 1$--2 mag in the first 100 days, while the $(V-R)_0$ and $(V-I)_0$ colors evolve slowly and become red by only $\sim 0.5$ mag. The $(B-V)_0$ color turns blue after t$\sim$120 d, when the nebular phase begins. The overall colors of \sn\ are bluer than those of SN 1999em at all phases, except in $V-I$ where these two objects have similar colors. This suggests that SN 2014cx has a higher photospheric temperature than SN 1999em.

\subsection{Bolometric Light Curve} \label{subsec:bololc}
To derive the bolometric luminosity of \sn\ , we first convert the extinction-corrected magnitudes in different bands to flux values at the effective wavelength, and then integrate the combined flux over wavelength. The luminosity can then be calculated using the integrated flux and a distance of 18 Mpc (see Section \ref{sec:obs}). During the first 50 days, when \emph{Swift} UV photometry was available, we integrated the UV flux and extrapolated it to the optical flux. As the UV flux decreases quickly, we assume that the UV contribution to the bolometric flux is marginal in the late plateau phase and negligible ($\lesssim 1$\%) during the nebular phase (\eg\ \citealt{2014ApJ...787..139D}). Owing to the lack of NIR data, we estimate the tail luminosity using Equation 3 in \citet{2010MNRAS.404..981M}, where a bolometric correction of $0.33 \pm 0.06$ mag is adopted.

\begin{figure}
\centering
\includegraphics[width=0.8\textwidth]{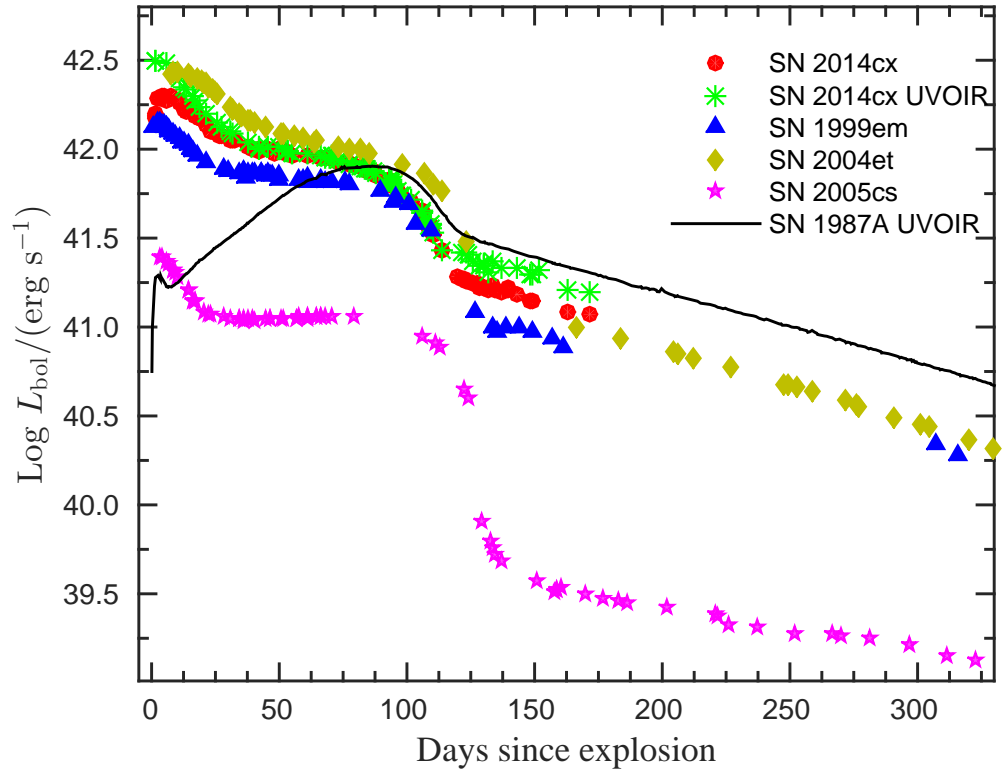}
\caption{Comparison of the bolometric luminosity evolution of \sn\ and other well-studied SNe IIP.}  \label{fig:bololc}
\end{figure}

Figure \ref{fig:bololc} shows the UV-optical-NIR (``UVOIR'') bolometric luminosity curve of \sn. One can see that it reached a peak of log [$L^{\rm peak}_{\rm bol}$/(\ergs)] = 42.47 at $\sim 3$ d after explosion. In Figure \ref{fig:bololc} we also compare the $UBVRI$ quasi-bolometric luminosity of \sn\ with that of some well-studied SNe IIP. It is readily seen that the luminosity evolution of \sn\ is similar to that of SN 2004et and SN 1999em but lies between these two objects in the early and plateau phases. In the nebular phase, we note that the tail luminosity of SN 2014cx is apparently higher than SN 1999em and even slightly higher than SN 2004et, suggesting that a relatively larger amount of nickel may be synthesized in its explosion (see discussion below).

\subsection{\nickel\ Mass} \label{subsec:nickmass}
For SNe IIP, the light curve in the nebular phase is powered mainly by the radioactive decay chain \nickel\ $\rightarrow$ \cobalt\ $\rightarrow$ \iron. And the tail luminosity is directly proportional to the mass of synthesized \nickel\, supposing that the gamma-ray deposition fraction is similar. For the well-studied type II-pec SN 1987A, the mass of \nickel\ has been determined to be $0.075 \pm 0.005$ \msun\ \citep{1996snih.book.....A}. For \sn\, we estimate the UVOIR bolometric luminosity at t $\approx$ 150 d to be 1.93 $\pm 0.14 \times 10^{41}$ \ergs by making a linear fit to the evolution between t $\approx$ 120 d and t $\approx$ 200 d. The luminosity of SN 1987A at the same phase is estimated to be 2.47 $\pm 0.02 \times 10^{41}$ \ergs. The ratio of \sn\ to SN 1987A is 0.78$\pm$0.06, which yields a value of $M_{\rm Ni} = 0.058 \pm 0.006$ \msun\ for \sn.

Based on the assumption that all $\gamma$-rays from \cobalt\ $\rightarrow$ \iron\ are entirely thermalized, \citet{2003ApJ...582..905H} found an independent relationship to estimate the value of \nickel\ mass using the tail luminosity ($L_{\rm t}$). We calculated $L_{\rm t}$ of \sn\ at 17 epochs from t $\approx$ 120 days to $\approx$ 170 days using the late-time $V$-band magnitude. The mean value of $M_{\rm Ni}$ resulting from Equation 2 of \citet{2003ApJ...582..905H} is $0.055 \pm 0.011$ \msun, consistent with the result from direct comparison with SN 1987A. 

Following the procedures of \citet{2003MNRAS.338..939E}, we also estimated the \nickel\ mass using the steepness parameter $S$, which is defined as the steepest value in the $V$-band light curve at the transitional phase from plateau to nebular tail. For \sn, we obtain $S = 0.071$ mag d$^{-1}$ and $M_{\rm Ni} = 0.056 \pm 0.007$ \msun, in good agreement with the values derived from the observed luminosity in the radioactive tail. The weighted mean value of $M_{\rm Ni}$ derived from above three results is $0.056 \pm 0.008$ \msun.

\section{Spectroscopic Analysis} \label{sec:spec}

\subsection{Optical Spectra} \label{subsec:spec}
\begin{figure}
\centering
\includegraphics[width=0.8\textwidth]{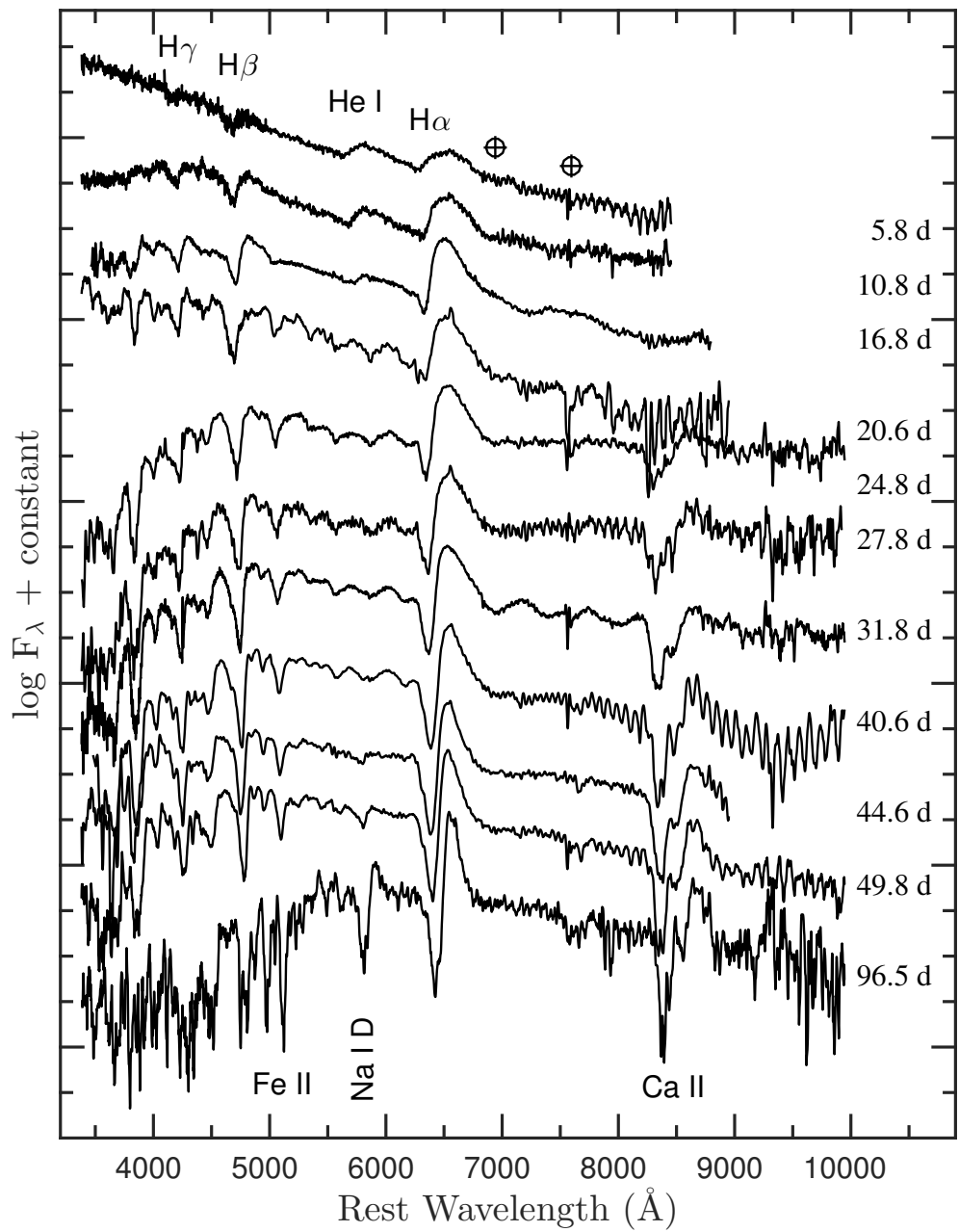}
\caption{Spectroscopic evolution of \sn\ during the photospheric phase. The spectra have been corrected for the redshift of the host galaxy \host\ ($z = 0.00549$, via NED). Some key features are labelled, as are residuals from telluric
absorption lines. The oscillations in some of the near-infrared spectra 
(most notably, at 40.6 d) are caused by incomplete removal of 
CCD fringing.} \label{fig:earlyspec}
\end{figure}

\begin{figure}
\centering
\includegraphics[width=0.8\textwidth]{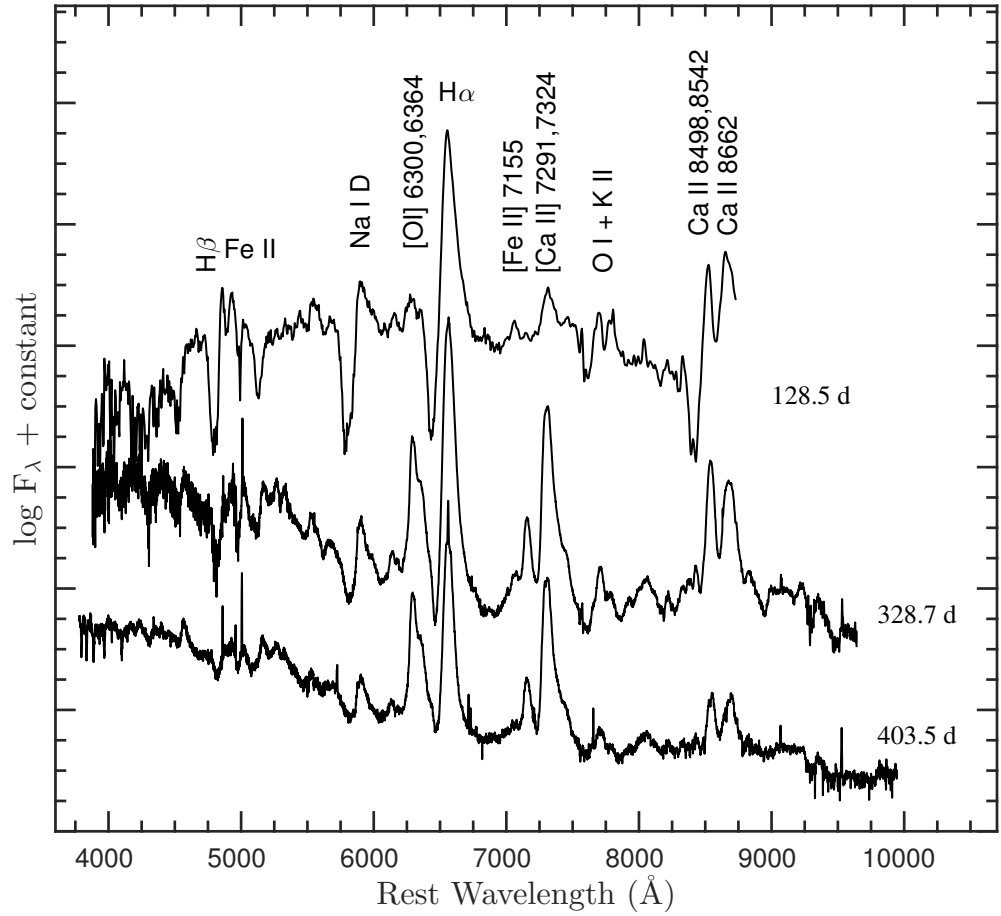}
\caption{Spectroscopic evolution of \sn\ during the nebular phase, when emission lines are dominant features.} \label{fig:latespec}
\end{figure}

The overall spectroscopic evolution of \sn\ is displayed in Figures \ref{fig:earlyspec} and \ref{fig:latespec}, covering early time ($\sim +6$ d) up to the nebular phase ($\sim +404$ d). All of the spectra have been corrected for the recession velocity of the host galaxy (1646 \kms) but not for the reddening. The main spectral features are labeled according to the lines previously identified for SNe IIP \citep{2002PASP..114...35L}.

At $t \approx 6$ days, the spectrum is very blue with a blackbody temperature exceeding $10^4$ K. The prominent features at such early phases are Balmer lines and He~I \ld5876 with broad P-Cygni profiles. At $t \approx 17$ days, the Fe~II \ld5169 absorption feature is visible, and it becomes stronger by $t \approx 21$ days along with Fe~II \ld5018. The Na~I and He~I lines are also detectable at $t \approx 21$ days, and both features grow stronger thereafter. The continuum becomes notably redder from $t \approx 21$ days to $t \approx 25$ days, suggesting a rapid decrease of the photospheric temperature during this period. Consequently, more metal lines (such as Sc~II, Ba~II, O~I, and the Ca~II NIR triplet) are formed, and they gradually become the dominant features in the spectra.

At $t \approx 128$ days, \sn\ starts to enter the nebular phase. The spectrum shows deep absorption of Na~I and prominent emission lines of [O~I] \ld\ld6300, 6364, [Fe~II] \ld\ld7155, and [Ca~II] \ld\ld7291, 7324. At $t \approx 329$ days and $t \approx 404$ days, the spectra exhibit emission lines of \ha, [O~I], [Fe~II], and [Ca~II], as well as P-Cygni profiles of Na~I, O~I, Fe~II, and the Ca~II NIR triplet. The multiple emission peaks at 5000--5500 \AA\ are likely contributed by [Fe~I] and [Fe~II] multiplets. We note that the continuum at 3800--5000 \AA\ is relatively bluer than that taken at 128 d, perhaps because of either a scattered light echo off the surrounding circumstellar material \citep[CSM;][]{2016MNRAS.457.3241A} or late-time CSM interaction \citep{2011MNRAS.417..261I}. For the light-echo scenario, the echoed light should come from earlier phases when the spectra were blue, and the scattering is more efficient at shorter wavelengths. However, the negligible extinction estimated for \sn\ in Section \ref {subsec:colorlc} implies that the CSM potentially causing the light echoes is in the opposite direction of \sn. In the case of CSM interaction, we can also see a hint of narrow \ha\ emission in the spectrum at t $\approx$ 404 days.

\begin{figure}
\centering
\includegraphics[width=0.8\textwidth]{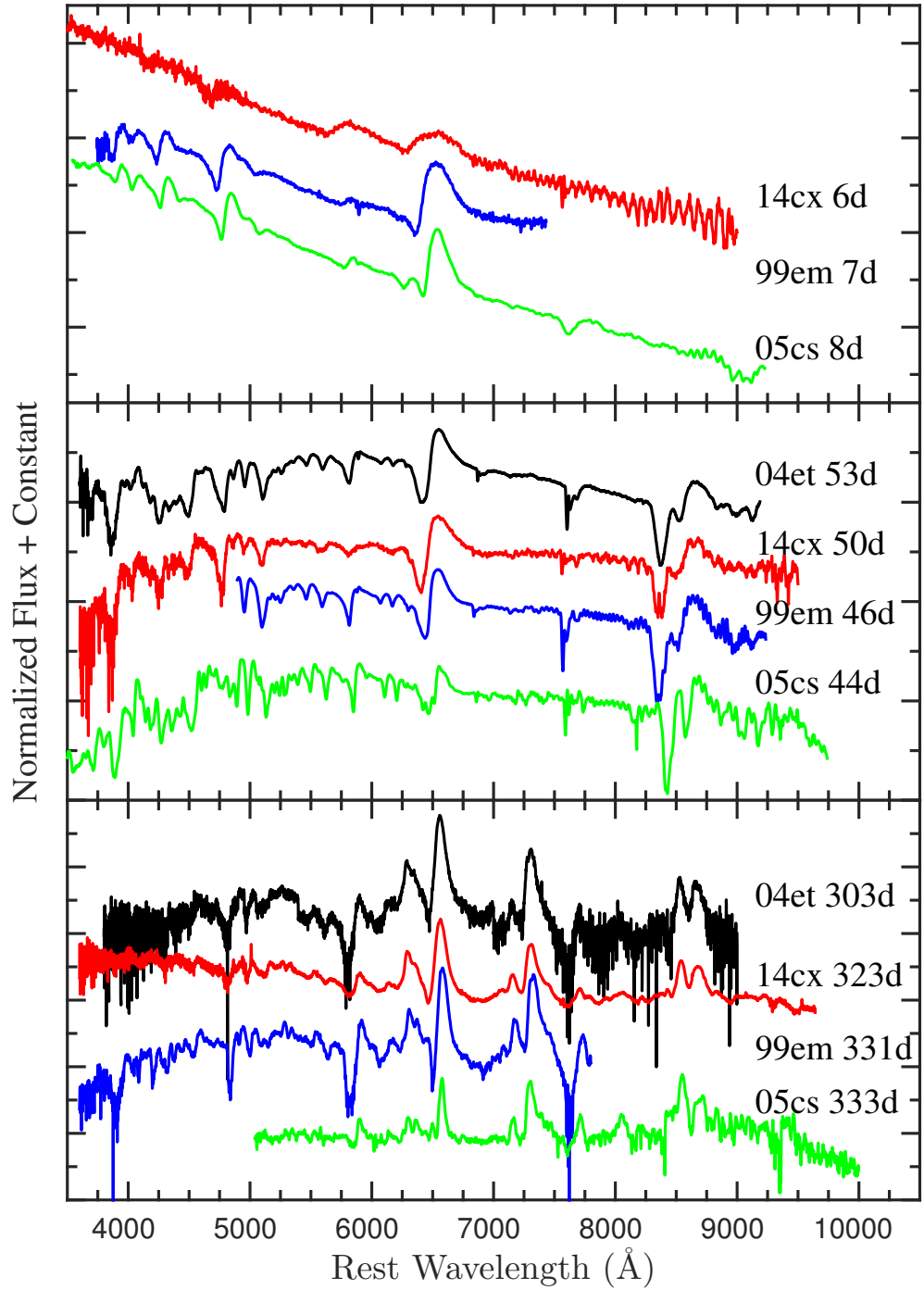}
\caption{Comparison of the reddening- and Doppler-corrected spectra of SNe 2014cx, 2004et, 1999em, and 2005cs at similar phases. Top panel, one week after explosion; middle panel, about 50 days after explosion; bottom panel, about one year after explosion.} \label{fig:spectra_compare}
\end{figure}

In Figure \ref{fig:spectra_compare}, we compare the spectra of \sn\ at about one week, 50 days, and one year after explosion with those of SN 1999em, SN 2005cs, and SN 2004et at similar phases. It can be seen that \sn\ has relatively shallower line profiles than the comparison SNe at early phases. During the plateau phase, $t \approx 50$ days after the explosion, the hydrogen lines of \sn\ are similar to those of the comparison SNe~IIP, but stronger than those of the subluminous SN 2005cs. In the nebular phase ($t \approx 300$ days), however, \sn\ appears to have weaker spectral features than SN 1999em and SN 2004et, but stronger than SN 2005cs, consistent with a flux contribution to the continuum.

\subsection{Expansion Velocities}
\begin{figure}
\centering
\includegraphics[width=0.8\textwidth]{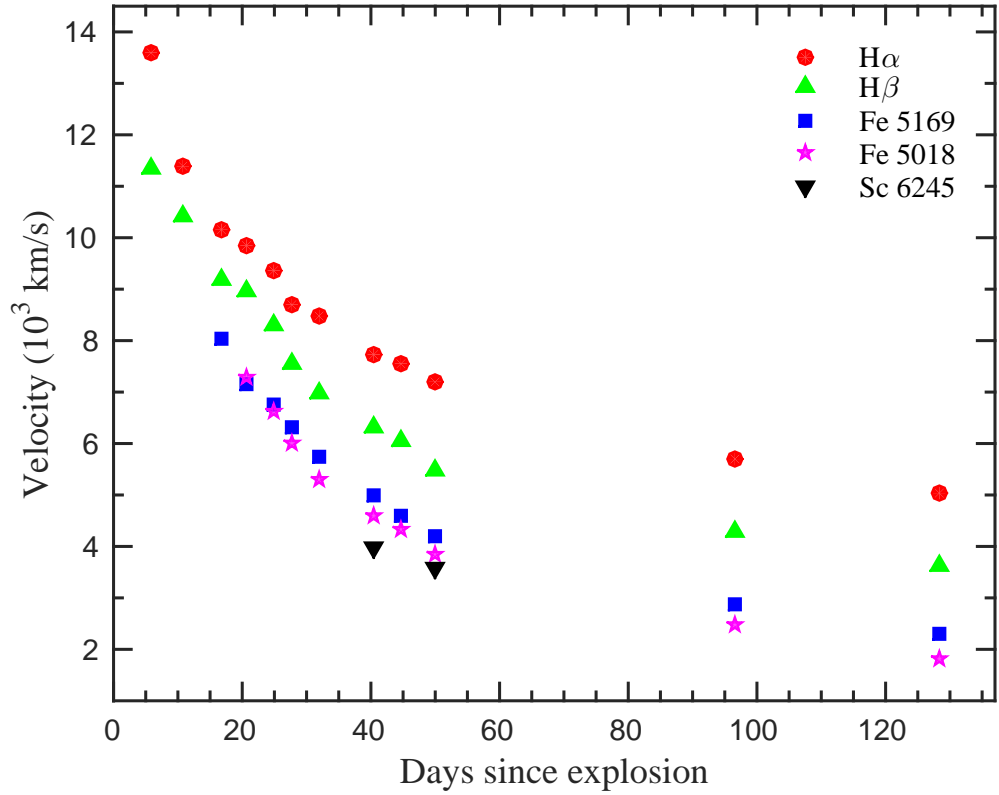}
\caption{The velocity evolution of \ha, \hb, and Fe II. The velocities are estimated using the Doppler shift of the absorption minima.}  \label{fig:vel}
\end{figure}

We measured the relativistic Doppler velocities of the \ha, \hb, Fe~II \ld5169, and Sc~II \ld6245 lines during the photospheric phase by fitting a Gaussian function to their absorption minima. The velocity evolution of these ions is shown in Figure \ref{fig:vel}. One can see that the velocities of hydrogen lines are higher than those of metal lines; the hydrogen lines have a lower optical depth and thus are formed at larger radii in the ejecta. The velocities seem to decline with an exponential trend.

\begin{figure}
\centering
\includegraphics[width=0.8\textwidth]{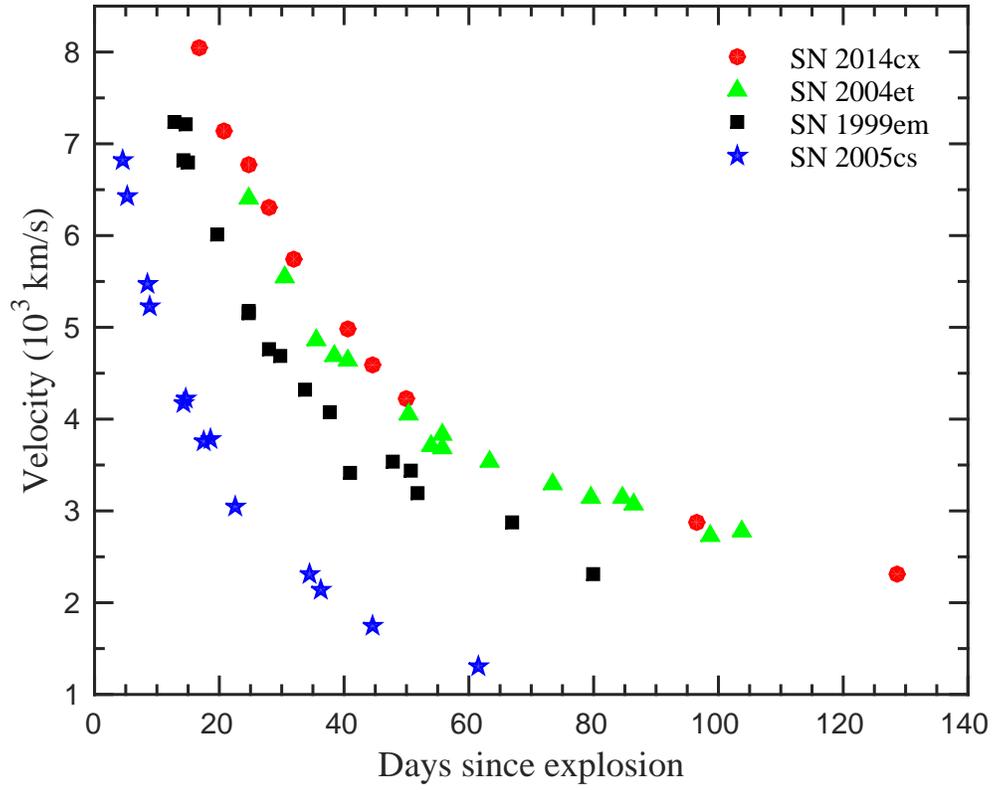}
\caption{Comparison of the expansion velocities of \sn\ measured from Fe II \ld5169 to those of the luminous Type IIP SN 2004et, the normal SN 1999em, and the subluminous SN 2005cs.}  \label{fig:vel_comp}
\end{figure}

To examine the differences of photospheric velocity between SN 2014cx and other SNe IIP, we compared the velocity as measured from Fe~II \ld5169 line. As shown in Figure \ref{fig:vel_comp}, \sn\ and SN 2014et have similar velocities at comparable phases, and their expansion velocities are higher than that of SN 1999em (by $\sim1000$ km s$^{-1}$) and SN 2005cs (by $\sim 3000$ km s$^{-1}$).

\section{Progenitor Estimates} \label{sec:model}
\subsection{Radius of Progenitor} \label{subsec:radi}
For CC~SNe, following the shock breakout, the heated stellar envelope expands and then cools down. The different timescales of photospheric temperature evolution depend mainly upon the initial radius of the progenitor and the opacity. A simple analytic function has been proposed by \citet{2011ApJ...728...63R} to estimate the radius of the progenitor of CC~SNe using the early-time temperature evolution. Theoretically, the duration that the photosphere can stay at a higher temperature depends on the radius of progenitor star, with longer time for a progenitor with a larger radius (and vice versa).

\begin{figure}
\centering
\includegraphics[width=0.8\textwidth]{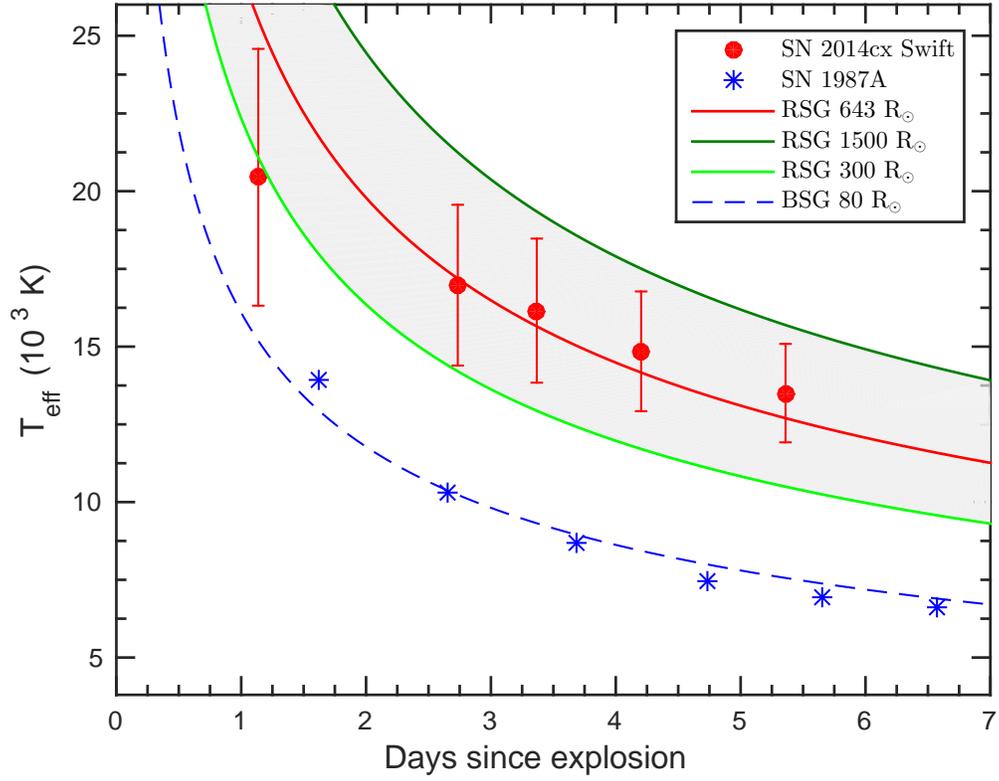}
\caption{Constraining the progenitor radius using the \citet{2011ApJ...728...63R} prescription. The red solid line is the best fit for a RSG of 643 \rsun, and the blue dashed line is for a BSG of 80 \rsun\ for SN 1987A.} \label{fig:radius}
\end{figure}

We constructed the spectral energy distribution (SED) and computed the blackbody temperatures using the \emph{Swift} UV and optical data obtained during the first week after explosion. The luminosity and temperature can then be used to constrain the radius of the progenitor star by using Equation 13 of \citet{2011ApJ...728...63R} \citep[e.g.,][]{2014MNRAS.438L.101V, 2015MNRAS.450.2373B, 2016ApJ...820...33R}. Note that the equation is only valid for the first week after explosion, when the light curve is dominated by shock cooling at very early phase, and the photosphere is located at the outer shell of the expanding ejecta \citep{2016ApJ...820...33R}. Adopting an optical opacity of 0.34 ${cm}^2 g^{-1}$ and a typical RSG density profile $f_{\rho}=0.13$, we obtained an initial radius of $643 \pm 60$ \rsun\ for the progenitor of \sn\ (see Figure \ref{fig:radius}), consistent with the typical size of an RSG.

Based on the SuperNova Explosion Code (SNEC, \citealt{2015ApJ...814...63M}), \citet{2016arXiv160308530M} suggests that the early-time evolution of the light curves of SNe IIP relies sensitively on the radius of the exploding star, as explained with an analytical correlation between the $g$-band rise time and the progenitor radius (i.e., log $R [R_{\odot}]$ = 1.225 log $t_{\rm rise}$ [day] + 1.692). Thus, we fit the early-time $g$-band light curve of SN 2014cx and obtain the rise time as $t_{rise}$= 7.90 $\pm$ 0.10 days. Inserting this value into the rise time-radius relation, we estimate the radius of the progenitor to be 619 $\pm$ 10 \rsun.

\subsection{Hydrodynamical Modeling} \label{subsec:model}
In this subsection, we further determine the main physical parameters of SN 2014cx and its progenitor (\ie\ the explosion energy, the radius of progenitor star, and the ejected mass) using a method of hydrodynamical modeling. This method uses the SN observables (\ie\ the bolometric light curve, the velocity evolution, and the temperature of the continuum obtained during the photospheric phase) as input parameters to constrain the physical properties of expanding ejecta and the evolution of SN observables (from the shock breakout up to the nebular phase) using the general-relativistic, radiation-hydrodynamics code. This technique has been successfully applied to the studies of numerous SNe IIP, including SNe 2007od, 2009bw, 2009E, 2012A, 2013ab, and 2013ej \citep{2011MNRAS.417..261I, 2012MNRAS.422.1122I, 2012A&A...537A.141P, 2013MNRAS.434.1636T, 2015MNRAS.450.2373B, 2015ApJ...807...59H}. Details about this hydro-dynamical model are well described in \citet{2003MNRAS.338..711Z, 2010MSAIS..14..123P, 2011ApJ...741...41P}. 

\begin{figure}
\centering
\includegraphics[width=0.8\textwidth]{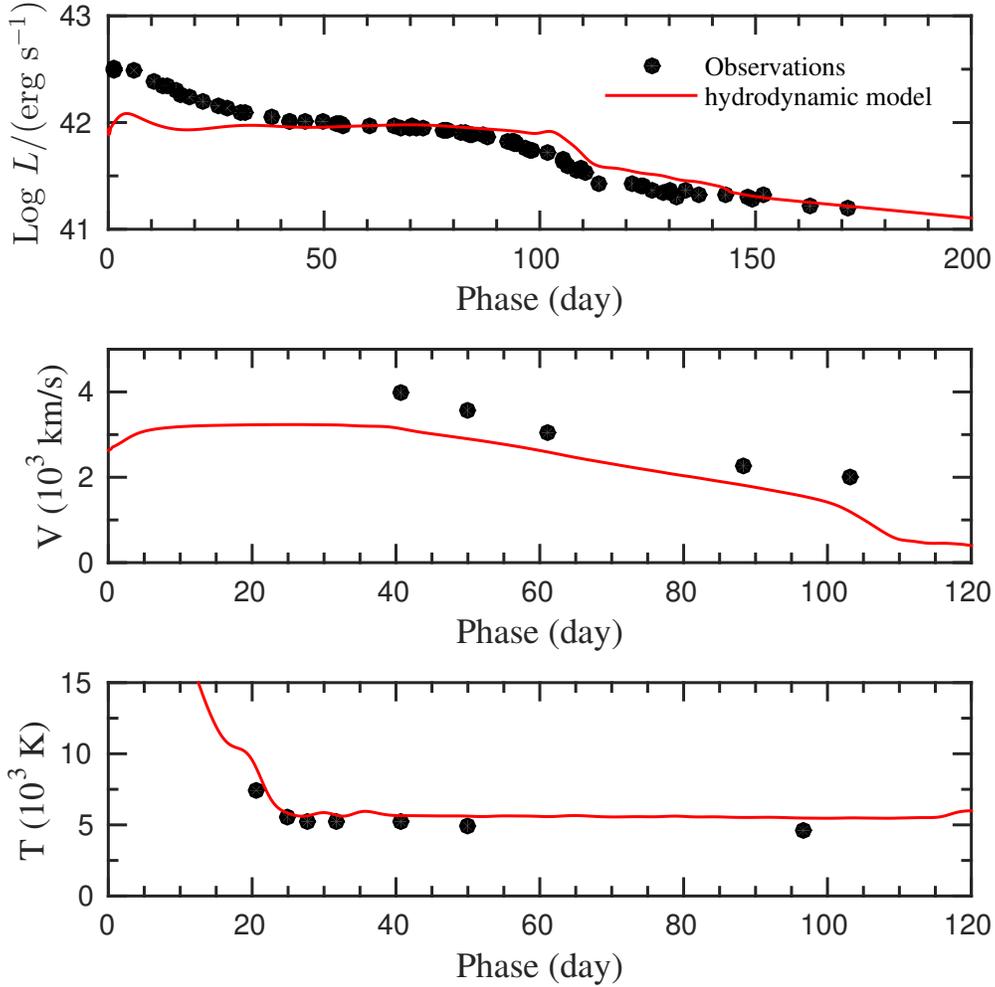}
\caption{Comparison of the evolution of the main observables of \sn\ with the best-fit model computed with the general-relativistic, radiation-hydrodynamics code (total energy $\sim 0.4 \times 10^{51}$ erg, initial radius $4 \times 10^{13}$ cm, envelope mass 8 \msun). Top, middle, and bottom panels show the bolometric light curve, the photospheric velocity, and the photospheric temperature as a function of time, respectively. To estimate the photospheric velocity from observations, we used the value inferred from the Fe II \ld5169 line.}  \label{fig:model}
\end{figure}

Based on the estimates of explosion epoch (MJD = 56,901.89; Section \ref{subsec:lc}), bolometric luminosity, and nickel mass (0.056 \msun; Section \ref{subsec:nickmass}), the best-fit hydrodynamic model returns a total (kinetic plus thermal) energy of $0.4 \times 10^{51}$ erg, an initial radius of $4 \times 10^{13}$ cm ($\sim 570$ \rsun), and an envelope mass of 8 \msun\ (see Figure \ref{fig:model}) for SN 2014cx. Considering a mass of $\sim 1.5-2.0$ \msun\ for the compact remnant star, we estimate that SN 2014cx has an immediate progenitor mass of 9.5--10.0 \msun\ when exploding. The zero-age main-sequence mass should be slightly higher given that the progenitor star suffers some mass loss during the lifetime. These values are consistent with those of a typical RSG with relatively low mass. The radius is also in agreement with that estimated from the early temperature as described in Section \ref{subsec:radi} within uncertainties.

We further compare our light curves and spectra with those from \citet{2009ApJ...703.2205K}, and we found that our parameters generally fall into the range of their Table 2, except that we have a smaller nickel mass (0.056 \msun\ vs. 0.1--0.5 \msun). This difference is likely due to that their numerical models use larger main sequence masses (12--25 \msun) than our hydrodynamic modeling results (9.5--10.0 \msun).

The evolutions of three observables (luminosity, velocity, and temperature) and the modeling results are shown in Figure \ref{fig:model}. As it can be seen, the modeling reasonably match the observed luminosity/temperature except at early phases where a larger deviation is seen in luminosity. This deviation is likely due to that the density profile in the radial direction of the outermost ejecta cannot be well produced by our simulations \citep{2011ApJ...741...41P}. In the middle panel of Figure \ref{fig:model}, we note that there is also a small ($\lesssim 10$--15\%) discrepancy between our best-fit model and the observed photospheric velocity. This discrepancy may be attributed to a systematic shift between the true photospheric velocity and the values estimated from the observed P-Cygni line profiles \citep{2005A&A...439..671D}. This can be explained with the fact that the optical depth is higher for spectral lines relative to the continuum, which can place the line photosphere at a larger radius (see also \citealt{2013A&A...555A.142I}).

\section{Discussion and Summary} \label{sec:sum}
In this paper, we present extensive UV and optical photometry and optical spectroscopy of \sn\ in \host, spanning the period from $-30$ d to $+404$ d from the maximum light. The explosion time is constrained to be MJD = 56,901.89 with an accuracy of $\pm 0.5$ day. 

The characteristics of the light curves, such as the rise time, duration of the plateau phase, post-peak decline, and bolometric luminosity, suggest that \sn\ is a normal type IIP supernova. The KAIT unfiltered and LCOGT r-band light curves seem to experience two brightening components, with the first likely related to shock breakout of the supernova. The plateau duration is $\sim 100$ days, similar to that of our comparison SNe~IIP. The value of $M_V$ at mid-plateau phase ($\sim 50$ d) is $-16.48 \pm 0.43$ mag for \sn, lying between the luminous SNe~IIP ($\sim -17$ mag, SN 2004et) and subluminous SNe~IIP ($\sim -15$ mag, SN 2005cs). The mass of \nickel\ using the tail luminosity and steepness methods yield a value of 0.056 \msun, similar to that of SN 1999em and SN 2004et.

The spectroscopic evolution of \sn\ shares a similarity with the typical Type IIP SNe 1999em and 2004et. The early-time spectra exhibit a nearly featureless continuum with only hydrogen Balmer lines and He~I visible. As the SN evolves, the continuum becomes redder and the metal lines emerge, becoming the dominant features during the photospheric phase. During the nebular phase, the spectra are dominated by strong emission lines. The continuum at 3800--5000 \AA\ is relatively blue, which might be caused by either late-time CSM interaction or a scattered-light echo. The value and evolution of the expansion velocity derived from Fe~II \ld5169 are similar to those of SN 2004et, but $\sim 1000$ km s$^{-1}$ higher than the expansion velocity of SN 1999em.

By modeling the observables of \sn\ as derived from our observations, we estimate that this explosion produces a total energy of $0.4 \times 10^{51}$ ergs and an ejected mass of $\sim$ 8.0 \msun. The progenitor star is calculated to have a radius of $4 \times 10^{13}$ cm ($\sim 574$ \rsun), which agrees well with that estimated from the early photospheric temperature evolution ($643 \pm 60$ \rsun) and $g$-band rise time -- radius relation from SNEC ($619 \pm 10$ \rsun). The values above are consistent with a core-collapse scenario from a typical RSG having an initial mass of 9.5--10 \msun.

\section*{Acknowledgments}
We thank the support of the staffs at Xinglong Station (National Astronomical Observatory of China), Li-Jiang Observatory (Yunnan Astronomical Observatory of China), and Lick Observatory for assistance with the observations. We thank Melina Bersten for discussion on early detection of shock-breakout. We also acknowledge the use of public data from the \emph{Swift} and Las Cumbres Observatory Global Telescope Network data archives. Some of the data presented herein were obtained at the W. M. Keck Observatory, which is operated as a scientific partnership among the California Institute of Technology, the University of California, and NASA; the observatory was made possible by the generous financial support of the W. M. Keck Foundation. KAIT and its ongoing operation were made possible by donations from Sun Microsystems, Inc., the Hewlett-Packard Company, AutoScope Corporation, Lick Observatory, the US National Science Foundation (NSF), the University of California, the Sylvia \& Jim Katzman Foundation, and the TABASGO Foundation. Research at Lick Observatory is partially supported by a generous gift from Google.

This work is supported by the Major State Basic Research Development Program (2013CB834903), the National Natural Science Foundation of China (NSFC grants 11178003, 11325313, and 11633002), and the Strategic Priority Research Program of Emergence of Cosmological Structures of the Chinese Academy of Sciences (grant No. XDB09000000). T.-M. Zhang is supported by the NSFC (grants 11203034). J.-J. Zhang is supported by the NSFC (grants 11403096), the Key Research Program of the CAS (Grant NO. KJZD-EW-M06) and the CAS ``Light of West China'' Program. DAH, CM, and GH are supported by NSF grant 1313484. A.V.F.'s group at U.C. Berkeley is grateful for financial assistance from NSF grant AST-1211916, the TABASGO Foundation, Gary and Cynthia Bengier, and the Christopher R. Redlich Fund. The work of A.V.F. was completed at the Aspen Center for Physics, which is supported by NSF grant PHY-1066293; he thanks the Center for its hospitality during the black holes workshop in June and July 2016.

\clearpage
\appendix

\clearpage
\newpage
\begin{table}[!ht]
\caption{Photometric Standard Stars in the Field of SN 2014cx ($1\sigma$ Uncertainties). } \label{tab:standstar}
\begin{tabular}{cccccccc}
\hline \hline
 Star& $\alpha_{\rm J2000}$&       $\delta_{\rm J2000}$& $U$ & $B$&  $V$&  $R$&  $I$\\
   ID&          (h m s)      &(\degr\,\arcmin\,\arcsec)  &(mag)&(mag)&(mag)&(mag)&(mag)\\
  \hline
1    &    1:00:07.08  &	 -7:30:33.54 &   20.07(09) & 19.35(06) & 18.31(04) & 17.69(05) & 17.06(09) \\
2    &    1:00:11.55  &	 -7:37:03.82 &	 18.61(04) & 18.39(05) & 17.56(03) & 17.08(04) & 16.59(07) \\
3    &    0:59:45.87  &	 -7:31:12.11 &	 20.17(11) & 19.71(06) & 18.66(04) & 18.04(04) & 17.43(08) \\
4    &    0:59:37.53  &	 -7:30:40.79 &	 21.32(25) & 20.43(08) & 18.79(05) & 17.69(11) & 16.26(22) \\
5    &    0:59:30.12  &	 -7:37:08.94 &	 18.66(10) & 18.56(03) & 18.20(02) & 17.98(03) & 17.77(05) \\
6    &    0:59:53.05  &	 -7:37:00.70 &	 18.26(04) & 18.24(04) & 17.52(03) & 17.08(03) & 16.65(06) \\
7    &    1:00:09.58  &	 -7:38:58.23 &	 18.38(04) & 18.07(05) & 17.19(03) & 16.66(04) & 16.12(08) \\
8    &    1:00:06.41  &	 -7:36:20.14 &	 20.47(15) & 19.52(08) & 17.96(04) & 17.01(07) & 16.03(15) \\
9    &    0:59:30.89  &	 -7:30:49.81 &	 19.23(06) & 18.56(05) & 17.58(04) & 17.00(04) & 16.44(08) \\
10   &    1:00:07.93  &	 -7:37:12.15 &	 19.69(08) & 19.38(05) & 18.45(03) & 17.89(04) & 17.31(08) \\
11   &    0:59:54.59  &	 -7:37:46.63 &	 18.04(04) & 17.44(05) & 16.47(03) & 15.89(04) & 15.34(08) \\
12   &    1:00:09.75  &	 -7:38:02.37 &	 17.27(03) & 17.11(05) & 16.28(03) & 15.77(04) & 15.24(08) \\
13   &    0:59:31.84  &	 -7:31:09.63 &	 20.16(10) & 19.10(08) & 17.56(05) & 16.62(07) & 15.61(15) \\
14   &    0:59:33.31  &	 -7:36:38.75 &	 18.08(09) & 17.86(04) & 17.12(03) & 16.69(03) & 16.28(06) \\
15   &    0:59:30.53  &	 -7:33:19.29 &	 16.08(03) & 17.22(04) & 17.29(03) & 17.37(04) & 17.50(08) \\
\hline
\end{tabular}
\end{table}

\newpage
\begin{longtable}{c c c c c c c c}
\caption{Optical Photometry of \sn\ from TNT ($1\sigma$ Uncertainties).} 
\label{tab:tnt} \\ 
\hline \hline
 UT Date &    MJD       & Phase$^{a}$  &  $U$    &     $B$     & $V$       & $R$    &    $I$ \\
(yy/mm/dd)&             & (day)       & (mag)   &    (mag)    &    (mag)   &   (mag)&   (mag)   \\
\hline
2014 Sep. 15 & 56915.75 &   13.86    & 14.30(04) & 15.23(03) & 14.89(04) & 14.70(10) & 14.60(05) \\  
2014 Sep. 17 & 56917.75 &   15.86    &  \nodata  & 15.14(03) & 14.89(07) & 14.70(07) & 14.56(03) \\  
2014 Oct. 12 & 56942.75 &   40.86    & 16.26(08  & 15.75(01) & \nodata   & \nodata   & \nodata   \\  
2014 Oct. 13 & 56943.75 &   41.86    & 16.37(12) & 15.87(04) & 15.08(04) & 14.75(06) & 14.46(04) \\  
2014 Oct. 21 & 56951.50 &   49.61    & 16.49(06) & 15.87(04) & 15.01(03) & 14.67(04) & 14.40(04) \\  
2014 Oct. 24 & 56954.50 &   52.61    & 16.57(11) & 15.95(03) & 15.05(02) & 14.78(04) & 14.39(04) \\  
2014 Oct. 25 & 56955.50 &   53.61    & 16.67(06) & 15.98(04) & 15.11(02) & 14.75(07) & \nodata   \\  
2014 Oct. 26 & 56956.50 &   54.61    & 16.79(05) & 16.07(02) & 15.11(03) & 14.73(05) & \nodata   \\  
2014 Nov. 08 & 56969.50 &   67.61    & 16.87(09) & 16.27(03) & 15.11(03) & 14.77(03) & 14.38(04) \\  
2014 Nov. 10 & 56971.50 &   69.61    & 17.07(08) & 16.27(05) & 15.03(03) & 14.75(04) & 14.44(05) \\  
2014 Nov. 11 & 56972.50 &   70.61    & 17.07(05) & 16.23(03) & 15.06(04) & 14.74(04) & 14.38(04) \\  
2014 Nov. 12 & 56973.50 &   71.61    &  \nodata  &   \nodata & 15.07(03) & 14.78(02) & \nodata   \\  
2014 Nov. 18 & 56979.50 &   77.61    & 17.28(16) &   \nodata & 15.18(05) & 14.89(07) & 14.40(06) \\  
2014 Nov. 19 & 56980.50 &   78.61    & 17.30(15) & 16.36(03) & 15.15(04) & 14.84(05) & 14.41(05) \\  
2014 Nov. 22 & 56983.50 &   81.61    & 17.46(06) & 16.42(05) & 15.19(05) & 14.82(06) & 14.50(04) \\  
2014 Nov. 23 & 56984.50 &   82.61    & 17.47(11) & 16.34(02) & 15.20(04) & 14.84(05) & 14.51(03) \\  
2014 Nov. 24 & 56985.50 &   83.61    &  \nodata  &   \nodata & 15.22(02) & 14.91(03) & 14.48(03) \\  
2014 Dec. 03 & 56994.50 &   92.61    &  \nodata  & 16.62(04) & 15.37(02) & 15.03(05) & 14.64(02) \\  
2014 Dec. 04 & 56995.50 &   93.61    &  \nodata  &   \nodata & 15.41(02) & 15.06(03) & 14.59(02) \\  
2014 Dec. 05 & 56996.50 &   94.61    & 17.79(25) & 16.73(04) & 15.45(02) & 15.12(03) & 14.65(06) \\  
2014 Dec. 07 & 56998.50 &   96.61    &  \nodata  & 16.82(05) & 15.66(02) & 15.18(02) & 14.61(04) \\  
2014 Dec. 08 & 56999.50 &   97.61    & 17.88(11) & 16.94(04) & 15.67(06) & 15.28(07) & 14.74(06) \\  
2014 Dec. 16 & 57007.50 &  105.61    &  \nodata  & 17.22(05) & 15.82(03) & 99.50(05) & 14.90(03) \\  
2014 Dec. 17 & 57008.50 &  106.61    & 18.03(09) & 17.52(06) & 16.13(04) & 15.46(06) & 15.04(04) \\  
2014 Dec. 19 & 57010.50 &  108.61    &  \nodata  & 17.27(05) & 16.21(04) & 15.64(05) & 15.23(05) \\  
2014 Dec. 20 & 57011.50 &  109.61    &  \nodata  & 17.57(07) & 16.10(06) & 15.51(05) & 15.22(07) \\  
2014 Dec. 21 & 57012.50 &  110.61    &  \nodata  & 17.58(05) & 16.28(03) & 15.68(04) & \nodata   \\  
2015 Jan. 01 & 57023.50 &  121.61    &  \nodata  & 18.04(09) & 17.01(06) & 16.30(03) & 15.92(04) \\  
2015 Jan. 03 & 57025.50 &  123.61    &  \nodata  & 18.20(06) & 17.02(07) & 16.31(05) & 15.86(03) \\  
2015 Jan. 06 & 57028.50 &  126.61    &  \nodata  &   \nodata &  \nodata  & 16.31(05) & 15.91(05) \\  
2015 Jan. 07 & 57029.50 &  127.61    &  \nodata  &   \nodata & 17.01(04) & 16.39(07) & 15.89(06) \\  
2015 Jan. 08 & 57030.50 &  128.61    &  \nodata  & 18.09(05) & 17.21(05) & 16.37(05) & 15.91(05) \\  
2015 Jan. 12 & 57034.50 &  132.61    &  \nodata  &   \nodata & \nodata   & 16.34(02) & 15.93(04) \\  
2015 Jan. 13 & 57035.50 &  133.61    &  \nodata  & 18.23(05) & 17.13(04) & 16.36(04) & 15.95(04) \\  
2015 Jan. 18 & 57040.50 &  138.61    &  \nodata  &   \nodata &  \nodata  & 16.42(03) & 16.00(02) \\ 
2015 Jan. 19 & 57041.50 &  139.61    &  \nodata  &   \nodata & 17.18(04) & 16.41(05) & 16.01(05) \\  
2015 Jan. 31 & 57053.50 &  151.61    &  \nodata  &   \nodata &  \nodata  & 16.67(06) & 16.20(04) \\ 
\hline
\end{longtable}
\begin{flushleft}
  $^{a}$Relative to the explosion date, MJD = 56,901.89.\\
   \end{flushleft}
 
   \newpage 
\begin{longtable}{cccccccc}
\caption{Optical Photometry of \sn\ from LCOGT ($1\sigma$ Uncertainties).} 
\label{tab:lcogt}\\
\hline \hline 
UT Date &    MJD       & Phase$^a$     &     $B$    & $V$       & $g$    &    $r$       & $i$    \\
(yy/mm/dd)&             & (day)       & (mag)   &    (mag)    &    (mag)   &   (mag)&   (mag)   \\
\hline
2014 Sep. 03 & 56903.130  &   1.240  & 15.32(03) & 15.32(01)     & 15.23(03) & 15.45(02) & 15.54(02) \\
2014 Sep. 03 & 56903.390  &   1.500  & 15.34(03) & 15.28(03)     & 15.16(04) & 15.36(03) & 15.51(02) \\
2014 Sep. 07 & 56907.770  &   5.880  & 15.12(03) & 15.01(03)     & 14.93(03) & 14.84(02) & 14.92(03) \\
2014 Sep. 12 & 56912.685  &  10.795  & 15.10(03) & 14.93(02)     & 14.96(01) & 14.98(03) & 14.98(02) \\
2014 Sep. 14 & 56914.755  &  12.865  & 15.20(03) & 14.92(03)     & 15.07(03) & 14.95(03) & 14.92(04) \\
2014 Sep. 16 & 56916.690  &  14.800  &\nodata    &\nodata        & 15.00(02) & 14.82(02) & 14.93(03) \\
2014 Sep. 18 & 56918.685  &  16.795  & 15.25(04) & 14.94(05)     & 15.03(03) & 14.81(02) & 14.92(07) \\
2014 Sep. 20 & 56920.545  &  18.655  & 15.30(02) & 14.91(02)     & \nodata   & 14.79(03) & 14.88(07) \\
2014 Sep. 23 & 56923.540  &  21.650  & 15.29(02) & 14.88(03)     & 15.04(04) & 14.76(03) & 14.94(05) \\
2014 Sep. 27 & 56927.545  &  25.655  & 15.45(03) & 14.92(02)     & 15.12(03) & 14.76(03) & 14.82(04) \\
2014 Sep. 29 & 56929.395  &  27.505  & 15.48(03) & 14.94(03)     & \nodata     &  \nodata    & \nodata \\
2014 Oct. 02 & 56932.520   &  30.630  & 15.58(03) & 14.94(02)     & 15.20(01) & 14.76(01) & 14.81(05) \\
2014 Oct. 03 & 56933.715   &  31.825  & 15.56(03) & 14.97(03)     & 15.29(04) & 14.78(02) & 14.79(03) \\
2014 Oct. 09 & 56939.690   &  37.800  & 15.73(04) & 15.01(03)     & 15.34(05) & 14.87(03) & 14.80(02) \\
2014 Oct. 13 & 56943.830   &  41.940  & 15.80(02) & 15.03(02)     & 15.39(03) & 14.80(02) & 14.85(02) \\
2014 Oct. 17 & 56947.475   &  45.585  & 15.86(03) & 15.06(02)     & 15.40(01) & 14.80(02) & 14.79(02) \\
2014 Oct. 21 & 56951.565   &  49.675  & \nodata   & \nodata       & 15.42(02) & 14.87(02) & 14.74(02) \\
2014 Oct. 25 & 56955.405   &  53.515 & 15.94(04) & 15.12(04)     & 15.47(03) & 14.76(04) & 14.80(03) \\
2014 Nov. 01 & 56962.565   &  60.675 & 16.06(03) & 15.07(03)     & 15.55(03) & 14.83(03) & 14.77(03) \\
2014 Nov. 06 & 56967.980   &  66.090 & 16.13(02) & 15.10(03)     & 15.54(03) & 14.82(03) & 14.79(03) \\
2014 Nov. 13 & 56974.630   &  72.740 & 16.26(03) & 15.16(03)     & 15.63(04) & 14.94(04) & 14.85(05) \\ 
2014 Nov. 18 & 56979.900   &  78.010 & 16.30(03) & 15.16(02)     & 15.68(03) & 14.89(02) & 14.88(03) \\
2014 Nov. 25 & 56986.060   &  84.170 & 16.43(03) & 15.26(02)     & 15.80(04) & 14.99(03) & 14.89(03) \\
2014 Nov. 27 & 56988.890   &  87.000 & 16.41(03) & 15.25(03)     & \nodata   &  \nodata  & \nodata \\
2014 Nov. 28 & 56989.780   &  87.890 & 16.49(04) & 15.32(03)     & 15.83(02) & 14.99(02) & 14.97(03) \\
2014 Dec. 04 & 56995.820   &  93.930 & 16.65(04) & 15.44(02)     & 15.96(04) & 15.08(03) & 15.00(02) \\
2014 Dec. 04 & 56995.850   &  93.960 & 16.71(04) & 15.47(03)     & \nodata   &  \nodata  & \nodata \\
2014 Dec. 09 & 57000.120   &  98.230 & 17.04(05) & 15.57(02)     & 16.15(04) & 15.33(03) & 15.11(04) \\
2014 Dec. 12 & 57003.830   & 101.940 & 16.95(04) & 15.67(02)     & 16.24(03) & 15.33(02) & 15.24(05) \\
2014 Dec. 16 & 57007.545   & 105.655 & 17.26(05) & 15.92(03)     & 16.56(04) & 15.51(03) & 15.46(02) \\
2014 Dec. 20 & 57011.480   & 109.590 & 17.63(04) & 16.25(02)     & 16.84(03) & 15.74(02) & 15.67(05) \\
2014 Dec. 24 & 57015.425   & 113.535 & 17.84(06) & 16.60(03)     & 17.18(04) & 16.03(01) & 15.92(03) \\
2014 Dec. 30 & 57021.430   & 119.540 & 18.09(07) & 17.04(04)     & 17.53(05) & 16.44(03) & 16.44(09) \\
2015 Jan. 02 & 57024.485   & 122.595 &\nodata    &\nodata        & 17.55(06) & 16.48(05) & 16.47(06) \\
2015 Jan. 03 & 57025.795   & 123.905 & 18.16(05) & 17.04(03)     & 17.59(04) & 16.48(03) & 16.43(05) \\
2015 Jan. 05 & 57027.830   & 125.940 & 18.22(05) & 17.14(03)     & 17.66(05) & 16.51(02) & 16.54(03) \\
2015 Jan. 08 & 57030.445   & 128.555 & 18.36(05) & 17.21(04)     & 17.76(04) & 16.59(03) & 16.58(03) \\
2015 Jan. 09 & 57031.815   & 129.925 & 18.33(05) & 17.19(03)     & \nodata   &  \nodata  & \nodata  \\
2015 Jan. 10 & 57032.060   & 130.170 & 18.31(05) & 17.15(03)     & \nodata   &  \nodata  & \nodata  \\
2015 Jan. 11 & 57033.795   & 131.905 & 18.37(05) & 17.28(02)     & 17.66(03) & 16.64(02) & 16.58(02) \\
2015 Jan. 14 & 57036.070   & 134.180 & 18.35(05) & 17.23(03)     & 17.69(04) & 16.63(03) & 16.61(04) \\
2015 Jan. 15 & 57037.445   & 135.555 &\nodata    &\nodata        & 17.81(04) & 16.58(03) & 16.62(03) \\
2015 Jan. 16 & 57038.790   & 136.900 & 18.40(07) & 17.23(04)     & 17.74(04) & 16.62(03) & 16.64(03) \\ \hline
2015 Jan. 22 & 57044.790   & 142.900 &\nodata    & 17.24(03)     & 17.76(04) & 16.71(05) & 16.65(05) \\ 
2015 Jan. 28 & 57050.085   & 148.195 & 18.44(07) & 17.32(04)     & 17.86(05) & 16.69(03) & 16.76(05) \\
2015 Jan. 29 & 57051.080   & 149.190 & 18.41(08) & 17.33(04)     & 17.93(06) & 16.72(03) & 16.82(05) \\ 
2015 Feb. 11 & 57064.775   & 162.885 & 18.49(08) & 17.52(04)     & 17.91(04) & 16.79(03) & 16.85(04) \\ 
2015 Feb. 20 & 57073.075   & 171.185 & 18.45(08) & 17.57(04)     & \nodata   &  \nodata  & \nodata  \\
2015 June 04 & 57177.420  & 275.530 & \nodata   & \nodata       & \nodata   & 17.81(04) & 18.06(10) \\
2015 June 22 & 57195.790  & 293.900 & \nodata   & \nodata       & \nodata   & 18.03(09) & 18.02(06) \\
2015 June 30 & 57203.395  & 301.505 & 18.92(05) & 18.44(05)     & 18.65(04) & 17.85(04) & 18.206(04) \\
2015 July 17 & 57220.260  & 318.370 & 18.93(05) & 18.52(04)     & 18.69(07) & 18.11(06) & 18.200(06) \\
2015 Aug.  13 & 57247.330  & 345.440 & \nodata   & \nodata       & 18.70(05) & 18.36(05) & 18.259(04) \\
2015 Aug.  29 & 57263.255  & 361.365 & \nodata   & \nodata  	 & 19.02(21) & 18.27(23) & \nodata \\
2015 Sep. 17 & 57282.390  & 380.500 & \nodata   & \nodata       & 19.03(05) & 18.50(05) & 18.198(05) \\
\hline
 \end{longtable}
\begin{flushleft}
  $^{a}$Relative to the explosion date, MJD = 56,901.89.\\
   \end{flushleft}

\newpage
\begin{table}[!ht]
\caption{Unfiltered Photometry of \sn\ ($1\sigma$ Uncertainties).} \label{tab:kait}
\begin{center}
\begin{tabular}{cccccccccc}
\hline \hline 
       MJD     &Phase$^a$      &     Mag        & Error      & Telescope &       MJD     &Phase$^{a}$& Mag         & Error      & Telescope   \\
    56871.49   &  -30.40   &  $>$ 18.76     &            &   KAIT    &    56949.38   & 47.49    &  14.87           &   0.08     &   KAIT    \\
    56876.49   &  -25.40   &  $>$ 19.16     &            &   KAIT    &    56950.31   & 48.42    &  14.73           &   0.12     &   KAIT    \\
    56877.48   &  -24.41   &  $>$ 19.06     &            &   KAIT    &    56952.31   & 50.42    &  14.78           &   0.06     &   KAIT    \\
    56879.41   &  -22.48   &  $>$ 18.50     &            &   KAIT    &    56954.33   & 52.44    &  14.79           &   0.05     &   KAIT    \\
    56880.41   &  -21.48   &  $>$ 18.67     &            &   KAIT    &    56957.31   & 55.42    &  14.86           &   0.05     &   KAIT    \\
    56881.39   &  -20.50   &  $>$ 18.38     &            &   KAIT    &    56958.33   & 56.44    &  14.84           &   0.05     &   KAIT    \\
    56886.46   &  -15.43   &  $>$ 19.48     &            &   KAIT    &    56959.28   & 57.39    &  14.80           &   0.07     &   KAIT    \\
    56887.47   &  -14.42   &  $>$ 19.05     &            &   KAIT    &    56960.26   & 58.37    &  14.88           &   0.06     &   KAIT    \\
    56888.45   &  -13.44   &  $>$ 19.22     &            &   KAIT    &    56964.29   & 62.40    &  14.90           &   0.07     &   KAIT    \\
    56889.45   &  -12.44   &  $>$ 19.14     &            &   KAIT    &    56970.32   & 68.43    &  14.80           &   0.09     &   KAIT    \\
    56890.45   &  -11.44   &  $>$ 19.13     &            &   KAIT    &    56971.26   & 69.37    &  14.86           &   0.07     &   KAIT    \\
    56891.44   &  -10.45   &  $>$ 19.14     &            &   KAIT    &    56972.26   & 70.37    &  14.80           &   0.06     &   KAIT    \\
    56892.47   &   -9.42   &  $>$ 19.26     &            &   KAIT    &    56976.32   & 74.43    &  14.90           &   0.07     &   KAIT    \\
    56893.41   &   -8.48   &  $>$ 19.15     &            &   KAIT    &    56977.25   & 75.36    &  14.85           &   0.06     &   KAIT    \\
    56894.48   &   -7.41   &  $>$ 18.88     &            &   KAIT    &    56978.25   & 76.36    &  14.85           &   0.12     &   KAIT    \\
    56895.42   &   -6.47   &  $>$ 19.17     &            &   KAIT    &    56985.23   & 83.34    &  14.89           &   0.08     &   KAIT    \\
    56896.41   &   -5.48   &  $>$ 19.22     &            &   KAIT    &    56986.24   & 84.35    &  14.96           &   0.06     &   KAIT    \\
    56897.49   &   -4.40   &  $>$ 19.14     &            &   KAIT    &    56987.23   & 85.34    &  15.09           &   0.05     &   KAIT    \\
    56898.40   &   -3.49   &  $>$ 19.13     &            &   KAIT    &    56988.26   & 86.37    &  15.11           &   0.08     &   KAIT    \\
    56900.40   &   -1.49   &  $>$ 19.19     &            &   KAIT    &    56989.19   & 87.30    &  14.99           &   0.06     &   KAIT    \\
    56901.39   &   -0.50   &  $>$ 19.11     &            &   KAIT    &    56990.18   & 88.29    &  15.07           &   0.05     &   KAIT    \\
    56902.40   &    0.51   &  15.69         &   0.05     &   KAIT    &    56999.24   & 97.35    &  15.27           &   0.07     &   KAIT    \\    
    56903.36   &    1.47   &  15.17         &   0.07     &   KAIT    &    57001.22   & 99.33    &  15.28           &   0.08     &   KAIT    \\
    56920.41   &   18.52   &  14.85         &   0.06     &   KAIT    &    57006.14   &104.25    &  15.47           &   0.09     &   KAIT    \\
    56922.33   &   20.44   &  14.65         &   0.06     &   KAIT    &    57015.12   &113.23    &  16.05           &   0.07     &   KAIT    \\
    56923.30   &   21.41   &  14.67         &   0.08     &   KAIT    &    57027.08   &125.19    &  16.54           &   0.15     &   KAIT    \\
    56924.30   &   22.41   &  14.65         &   0.05     &   KAIT    &    56902.57   &  0.68    &  15.6            &            &   Koichi Itagaki  \\
    56944.36   &   42.47   &  14.70         &   0.07     &   KAIT    &    56903.04   &  1.15    &  15.2            &            &   T. Yusa     \\
    56946.35   &   44.46   &  14.83         &   0.11     &   KAIT    &    56906.93   &  5.04    &  14.8            &            &   Paolo Campaner   \\ 
\hline 
\end{tabular}
\begin{flushleft}
  $^{a}$Relative to the explosion date, MJD = 56,901.89.\\
   \end{flushleft}
\end{center}
\end{table}

\newpage
 \begin{table}
  \caption{UV and Optical Photometry of \sn\ from $Swift$ ($1\sigma$ Uncertainties).}    \label{tab:uvot}
   \begin{tabular}{ccccccccc}
   \hline \hline
   UT Date &    MJD       & Phase$^a$       & $uvw$2       &    $uvm$2   &    $uvw$1  &   $U$ &     $B$    & $V$   \\
   (yy/mm/dd)&            & (day)       & (mag)        &    (mag)    &    (mag)   &   (mag)&   (mag)   & (mag) \\
  \hline 
 2014 Sep. 3 & 56903.03 &  1.14   & 13.62(05)  & 13.70(05)  & 13.80(05)  & 13.95(04) &  15.34(06)  & 15.42(07) \\ 
 2014 Sep. 4 & 56904.62 &  2.73   & 13.70(05)  & 13.67(05)  & 13.66(05)  & 13.74(04) &  14.97(05)  & 15.09(06) \\
 2014 Sep. 5 & 56905.26 &  3.37   & 13.86(05)  & 13.71(05)  & 13.77(05)  & 13.78(04) &  14.95(05)  & 15.09(06) \\
 2014 Sep. 6 & 56906.09 &  4.20   & 14.06(05)  & 13.85(05)  & 13.80(05)  & 13.78(04) &  14.96(05)  & 15.01(06) \\
 2014 Sep. 7 & 56907.25 &  5.36   & 14.32(06)  & 14.03(06)  & 13.92(05)  & 13.78(04) &  14.92(05)  & 15.00(06) \\
 2014 Sep. 8 & 56908.31 &  6.42   & 14.54(07)  & \nodata    & 14.00(05)  & 13.80(04) &  14.88(05)  & \nodata   \\  
 2014 Sep. 9 & 56909.09 &  7.20   & 14.80(07)  & 14.40(06)  & 14.11(05)  & 13.83(04) &  14.92(05)  & 14.97(06) \\  
 2014 Sep. 12& 56912.91 &  11.02  & 15.48(09)  & 15.19(08)  & 14.60(06)  & 13.99(04) &  14.90(05)  & 14.97(06) \\ 
 2014 Sep. 16& 56916.28 &  14.39  & 16.22(11)  & 16.25(10)  & 15.30(08)  & 14.24(05) &  15.02(05)  & 14.99(06) \\ 
 2014 Sep. 20& 56920.78 &  18.89  & 17.67(24)  & 17.80(24)  & 16.32(11)  & 14.62(05) &  15.25(06)  & 14.96(06) \\ 
 2014 Sep. 24& 56924.71 &  22.82  & 18.21(36)  & 18.37(42)  & 16.82(15)  & 15.28(07) &  15.35(06)  & 14.94(06) \\ 
 2014 Sep. 28& 56928.60 &  26.71  & 18.55(46)  & \nodata    & 17.51(21)  & 15.62(07) &  15.45(06)  & 14.96(06) \\ 
 2014 Oct.  4 & 56934.23 &  32.34  & \nodata    & \nodata    & 18.08(30) &  15.85(07)  & 15.62(06)  & 14.95(07) \\ 
 2014 Oct. 11 & 56941.27 &  39.38  & \nodata    & \nodata    & \nodata   &  16.22(07)  & 15.80(06)  & 15.08(06) \\ 
 2014 Oct. 19 & 56949.91 &  48.02  & \nodata    & \nodata    & 18.62(47) &  16.46(07)  & 16.00(06)  & 15.12(06) \\ 
 2014 Oct. 22 & 56952.70 &  50.81  & \nodata    & \nodata    & \nodata   &  16.39(14)  & 15.96(09)  & 15.06(09) \\ 
 2014 Oct. 24 & 56954.60 &  52.71  & \nodata    & \nodata    & \nodata   &  16.56(07)  & 16.01(06)  & 15.09(06) \\ 
  \hline
   \end{tabular}
\begin{flushleft}
  $^{a}$Relative to the explosion date, MJD = 56,901.89.\\
   \end{flushleft}
  \end{table}

\newpage
\begin{table}
\caption{Photometric Parameters of SN 2014cx.}\label{tab:photopara}
\begin{center}
\begin{tabular}{cccccc}
\hline \hline 
                 &    $U$   & $B$   & $V$   & $R$   & $I$   \\
Peak magnitude   &    13.75 & 15.08 & 14.91 & --    & --    \\
Phase of maximum$^a$        &  3.29 &  8.24 & 10.83 & --    & --    \\  
Plateau magnitude&    --    &  --   & 15.10 & 14.75 & 14.38  \\
Decay rate (mag/100 d)      &   --  &  0.42 & 0.95  & 1.09 & 1.13 \\ 
\hline
\end{tabular}
\end{center}
\begin{flushleft}
  $^{a}$Relative to the explosion date, MJD = 56,901.89.\\
   \end{flushleft}
\end{table}

\newpage
\begin{table}[!htp]
\caption{Observing Log for Optical Spectra of \sn.} \label{tab:spelog}
\begin{center}
\begin{tabular}{lccccc}
\hline \hline
     UT Date        &   MJD        &    Phase$^a$       & Range        & Exposure & Telescope + Instrument\\
             &              &    (days)                 & (\AA)        & (s)      &          \\
   \hline
   2014 Sep. 7   &  56907.74    &     5.8                   & 3400--8500   & 2700     & LCOGT 2.0~m Telescope South + FLOYDS \\
   2014 Sep. 12  &  56912.69    &    10.8                   & 3400--8500   & 1800     & LCOGT 2.0~m Telescope South + FLOYDS \\
   2014 Sep. 18  &  56918.66    &    16.8                   & 3480--8850   & 2100     & Xinglong 2.16~m + BFOSC \\
   2014 Sep. 22  &  56922.46    &    20.6                   & 3400--9000   & 1800     & LCOGT 2.0~m Telescope South + FLOYDS \\
   2014 Sep. 26  &  56926.71    &    24.8                   & 3400--10,000  & 1800    & LCOGT 2.0~m Telescope South + FLOYDS \\
   2014 Sep. 29  &  56929.65    &    27.8                   & 3400--10,000   & 1800   & LCOGT 2.0~m Telescope South + FLOYDS \\
   2014 Oct.  3  &  56933.70    &    31.8                   & 3400--10,000   & 2700   & LCOGT 2.0~m Telescope South + FLOYDS \\
   2014 Oct. 12  &  56942.46    &    40.6                   & 3400--10,000   & 2700   & LCOGT 2.0~m Telescope South + FLOYDS \\
   2014 Oct.  16  &  56946.60    &    44.6                   & 3500--9000   & 2100    & Lijiang 2.4~m + YFOSC \\
   2014 Oct.  21  &  56951.65    &    49.8                   & 3400--10,000   & 2700  & LCOGT 2.0~m Telescope South + FLOYDS \\
   2014 Dec.   7  &  56998.36    &    96.5                   & 3400--10,000   & 2700  & LCOGT 2.0~m Telescope North + FLOYDS \\
   2015 Jan.   8  &  57030.43    &    128.6                  & 3900--8800   & 2400     & Xinglong 2.16~m + BFOSC \\
   2015 July  27  &  57230.62    &    328.7                  & 3900--9700   &  900     & Gemini-North 8.1~m + GMOS\\
   2015 Oct.  10  &  57305.38    &    403.5                  & 3800--10,000  & 1200     & Keck-I 10~m + LRIS \\
   \hline
 \end{tabular}
 \end{center}
\begin{flushleft}
  $^{a}$Relative to the explosion date, MJD = 56,901.89.\\
   \end{flushleft}
\end{table}

\end{document}